\documentclass[onecolumn, superscriptaddress, showpacs,
 amsmath,amssymb,
 aps,
 prx
]{revtex4-2}
\usepackage{graphicx}
\usepackage{xcolor}

\begin{document}

\title{Imaging with GeV muons produced via laser-wakefield-accelerated electrons}

\author{M. Dobre}
\email{These authors contributed equally to this work.}
\affiliation{Department of Nuclear Physics, "Horia Hulubei" National Institute for R\&D in Physics and Nuclear Engineering, Strada Reactorului 30, Măgurele, 077125, Romania}

\author{P. Ghenuche}
\email{These authors contributed equally to this work.}
\affiliation{Extreme Light Infrastructure - Nuclear Physics, "Horia Hulubei" National Institute for R\&D in Physics and Nuclear Engineering, Strada Reactorului 30, Măgurele, 077125, Romania}

\author{A. B\u{a}l\u{a}ceanu}
\affiliation{Extreme Light Infrastructure - Nuclear Physics, "Horia Hulubei" National Institute for R\&D in Physics and Nuclear Engineering, Strada Reactorului 30, Măgurele, 077125, Romania}

\author{D. Catana}
\affiliation{Extreme Light Infrastructure - Nuclear Physics, "Horia Hulubei" National Institute for R\&D in Physics and Nuclear Engineering, Strada Reactorului 30, Măgurele, 077125, Romania}
\affiliation{Faculty of Physics, University of Bucharest, 
Strada Atomiștilor 405, Măgurele, 077125, Romania}

\author{M.O. Cernaianu}
\affiliation{Extreme Light Infrastructure - Nuclear Physics, "Horia Hulubei" National Institute for R\&D in Physics and Nuclear Engineering, Strada Reactorului 30, Măgurele, 077125, Romania}

\author{D. Doroban\textcommabelow{t}u}
\affiliation{Extreme Light Infrastructure - Nuclear Physics, "Horia Hulubei" National Institute for R\&D in Physics and Nuclear Engineering, Strada Reactorului 30, Măgurele, 077125, Romania}
\affiliation{Faculty of Physics, University of Bucharest, 
Strada Atomiștilor 405, Măgurele, 077125, Romania}

\author{R. Lica}
\affiliation{Department of Nuclear Physics, "Horia Hulubei" National Institute for R\&D in Physics and Nuclear Engineering, Strada Reactorului 30, Măgurele, 077125, Romania}

\author{V. Malka}
\affiliation{Extreme Light Infrastructure - Nuclear Physics, "Horia Hulubei" National Institute for R\&D in Physics and Nuclear Engineering, Strada Reactorului 30, Măgurele, 077125, Romania}
\affiliation{Department of Physics of Complex Systems, Weizmann Institute of Science, Rehovot 7610001, Israel}

\author{D. Martello}
\affiliation{Dipartimento di Matematica e Fisica ”E. De Giorgi” dell’Università del Salento and Sezione INFN,  Via per Arnesano, 73100, Lecce, Italy}

\author{I. Mitu}
\affiliation{Extreme Light Infrastructure - Nuclear Physics, "Horia Hulubei" National Institute for R\&D in Physics and Nuclear Engineering, Strada Reactorului 30, Măgurele, 077125, Romania}

\author{M. Niculescu-Oglinzanu }
\affiliation{Department of Nuclear Physics, "Horia Hulubei" National Institute for R\&D in Physics and Nuclear Engineering, Strada Reactorului 30, Măgurele, 077125, Romania}

\author{L. Stan}
\affiliation{Department of Nuclear Physics, "Horia Hulubei" National Institute for R\&D in Physics and Nuclear Engineering, Strada Reactorului 30, Măgurele, 077125, Romania}

\author{D. Stanca}
\affiliation{Department of Nuclear Physics, "Horia Hulubei" National Institute for R\&D in Physics and Nuclear Engineering, Strada Reactorului 30, Măgurele, 077125, Romania}

\author{P. Tomassini }
\affiliation{Extreme Light Infrastructure - Nuclear Physics, "Horia Hulubei" National Institute for R\&D in Physics and Nuclear Engineering, Strada Reactorului 30, Măgurele, 077125, Romania}

\author{C.A. Ur}
\affiliation{Extreme Light Infrastructure - Nuclear Physics, "Horia Hulubei" National Institute for R\&D in Physics and Nuclear Engineering, Strada Reactorului 30, Măgurele, 077125, Romania}

\author{C. Vancea}
\affiliation{Department of Nuclear Physics, "Horia Hulubei" National Institute for R\&D in Physics and Nuclear Engineering, Strada Reactorului 30, Măgurele, 077125, Romania}

\author{A. Saftoiu}
\email{Corresponding author: alexandra.saftoiu@nipne.ro}  
\affiliation{Department of Nuclear Physics, "Horia Hulubei" National Institute for R\&D in Physics and Nuclear Engineering, Strada Reactorului 30, Măgurele, 077125, Romania}

\author{D. Doria}
\email{Corresponding author: domenico.doria@eli-np.ro}
\affiliation{Extreme Light Infrastructure - Nuclear Physics, "Horia Hulubei" National Institute for R\&D in Physics and Nuclear Engineering, Strada Reactorului 30, Măgurele, 077125, Romania}

\begin{abstract}

Artificially generated muon beams, with tunable parameters, are highly valuable in imaging objects that are opaque to any other scanning technique. 
Lower-energy muon beams are currently produced at accelerator facilities, while higher energies have only recently been observed as a result of the interaction of laser-wakefield-accelerated electrons and solid targets.
This work focuses on using muography imaging detectors to measure the profile of a muon beam generated via the Bethe-Heitler process and to perform muon imaging in an experimental setup extending up to 42 meters from the muon source. By comparing the data with Monte Carlo simulations, we confirm that the measured beam profile and object imaging are consistent with artificially produced muons of energy of a few GeV. To our knowledge, this represents the first demonstration of imaging dominated by an artificial, laser-driven muon beam.

\end{abstract}


\maketitle

\section{Introduction}\label{sec1}

Since its discovery, the muon, the second-generation lepton, has been the focus of numerous fundamental physics experiments from inferring the properties of the most energetic particles in the Universe to testing the Standard Model.

Because of their large mass relative to electrons, muons have reduced radiative energy loss and can, therefore, traverse considerable thicknesses of dense matter \cite{Woodley:2024}. This led to the emergence of muography, the scanning technique relying on muons to "see" inside inaccessible volumes, an inherently non-invasive tool applicable in various fields of research where the volumes of interest are either large, \textit{e.g.,} archaeology \cite{LAlvarez:1970, Morishima:2017}, geology \cite{Tanaka:2007, LoPresti:2020, Schouten:2018}, civil engineering \cite{Cimmino:2019, Borselli:2022, Thompson:2020}, geoscience \cite{Ariga:2018, Olah:2024}, or dense and requiring non-invasive scanning, \textit{e.g.,} nuclear safety, fuel deposits etc \cite{Pesente:2009, Russo:2014, Bouteille:2016, Baesso:2014}.

\par 
So far, muography has relied on the naturally available cosmic muons produced, alongside other secondaries, by the interaction of primary cosmic particles with the nuclei in the Earth's atmosphere.
However, the cosmic muon flux is relatively low, at 1 particle/cm$^2$/s, which leads to long exposure times. Moreover, cosmic muons have a zenith-centered angular distribution and a steep energy dependence~\cite{Grieder2001}.

An artificially produced muon beam would bypass many of the shortcomings of the natural flux and provide better control over the source parameters. Laser Wakefield Acceleration\cite{Tajima:1979,Esarey:2009} has been proven to create electrons in the 10 GeV range \cite{Gonsalves:2019,Picksley:2024,Rockafellow:2025}, which, when hitting a solid target, produce a large number of secondary particles, including muons \cite{Zhang:2025,Terzani:2025}.

In addition to muography applications, an artificial muon source is also interesting for muon spin spectroscopy ($\mu$SR) \cite{Nagamine:2008}, or studies of muon and electron neutrino oscillations ($\mu^+ \rightarrow e^+ + \nu_e + \bar{\nu_{\mu}}, \mu^- \rightarrow e^- + \bar{\nu_e} + \nu_{\mu}$) \cite{Bulanov:2004, Pakhomov:2002, Achenbach:2025}.  

Muon production is driven by several distinct mechanisms, such as: 1) $\pi^{\pm}$/K$^{\pm}$ decays ($\pi^{+(-)} \rightarrow \mu^{+(-)} + \nu_{\mu}(\bar{\nu_{\mu}})$, $K^{+(-)} \rightarrow \mu^{+(-)} + \nu_{\mu}(\bar{\nu_{\mu}})$), with a dominant contribution to the muon yield by the pions due to the energy production constraints and branching rations,
2) the Bethe-Heitler di-muon process ($e^- + A \rightarrow e^{-*} + A + \gamma,  \gamma + A \rightarrow \mu^+ + \mu^- + A$), whereby secondary bremsstrahlung gamma photons interact with the target nuclei to create muon-antimuon pairs \cite{rao:2018, Titov:2009}, and 3) the di-muon electro-production ($e^- + A  \rightarrow e^{-*} + A + \mu^+ + \mu^-$) \cite{Titov:2009, Yu:2024}. At higher energies, other di-muon processes ($e^- + \gamma \rightarrow e^{-*} + \mu^+ + \mu^-, \gamma + \gamma \rightarrow \mu^+ + \mu^- $) can contribute to the muon yield, but their cross-sections are significantly smaller at the experimental energies reached in this work.
\par Muon production from laser-wakefield-accelerated electrons was confirmed in 2025, at the Shanghai Superintense Ultrafast Laser Facility (SULF), where primary electrons with energies up to 1.5 GeV were available \cite{Zhang:2025}, and at the BELLA PW laser, using electrons with energies up to $\sim$8 GeV \cite{Terzani:2025}. Both measurements confirmed the presence of muons by recording the electrons resulting from muon decays.
This work reports on measurements of muon beam profile and transmission imaging of an object, validating the presence of muons by comparison with detailed Monte Carlo simulations based on the measured electron spectra.

\section{The experimental set-up}\label{setup}

\begin{figure}[ht]
\centering
\includegraphics[width=0.99\linewidth]{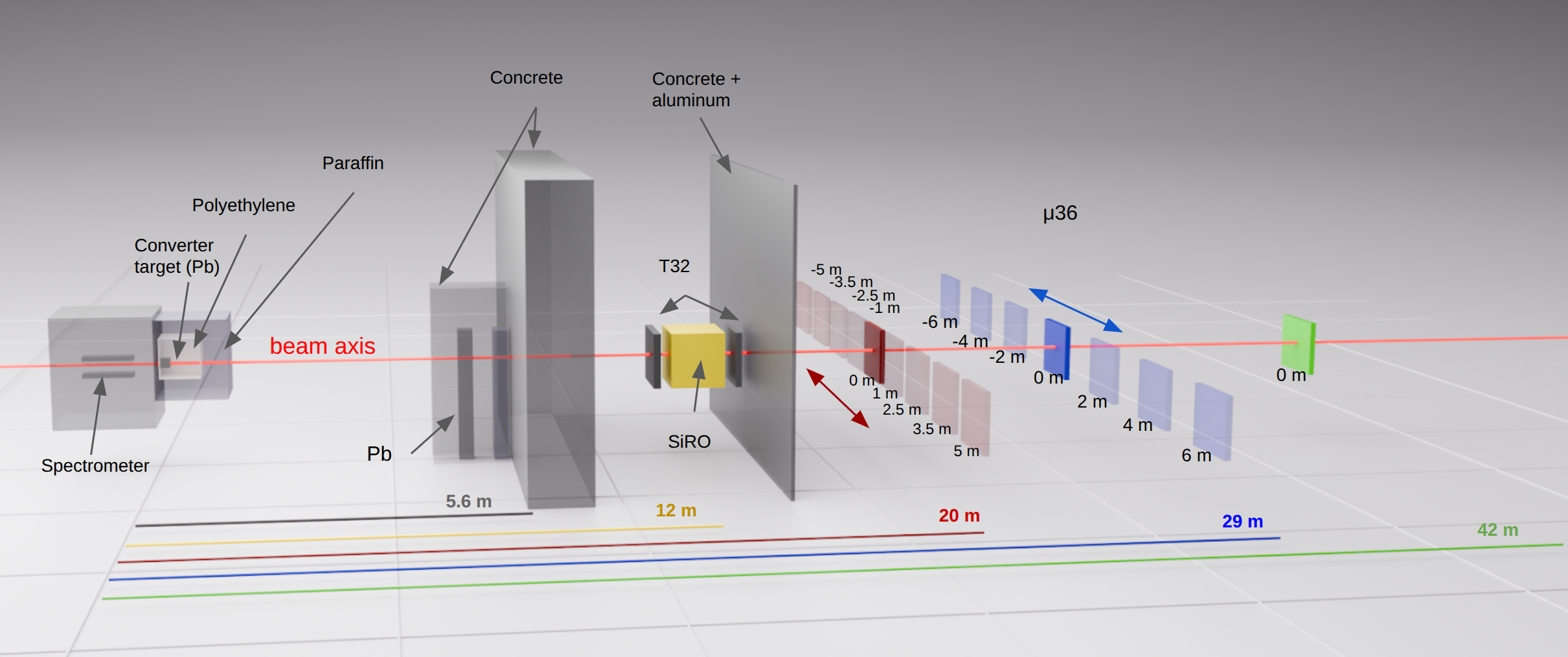}\\
\reflectbox{\includegraphics[width=0.43\linewidth]{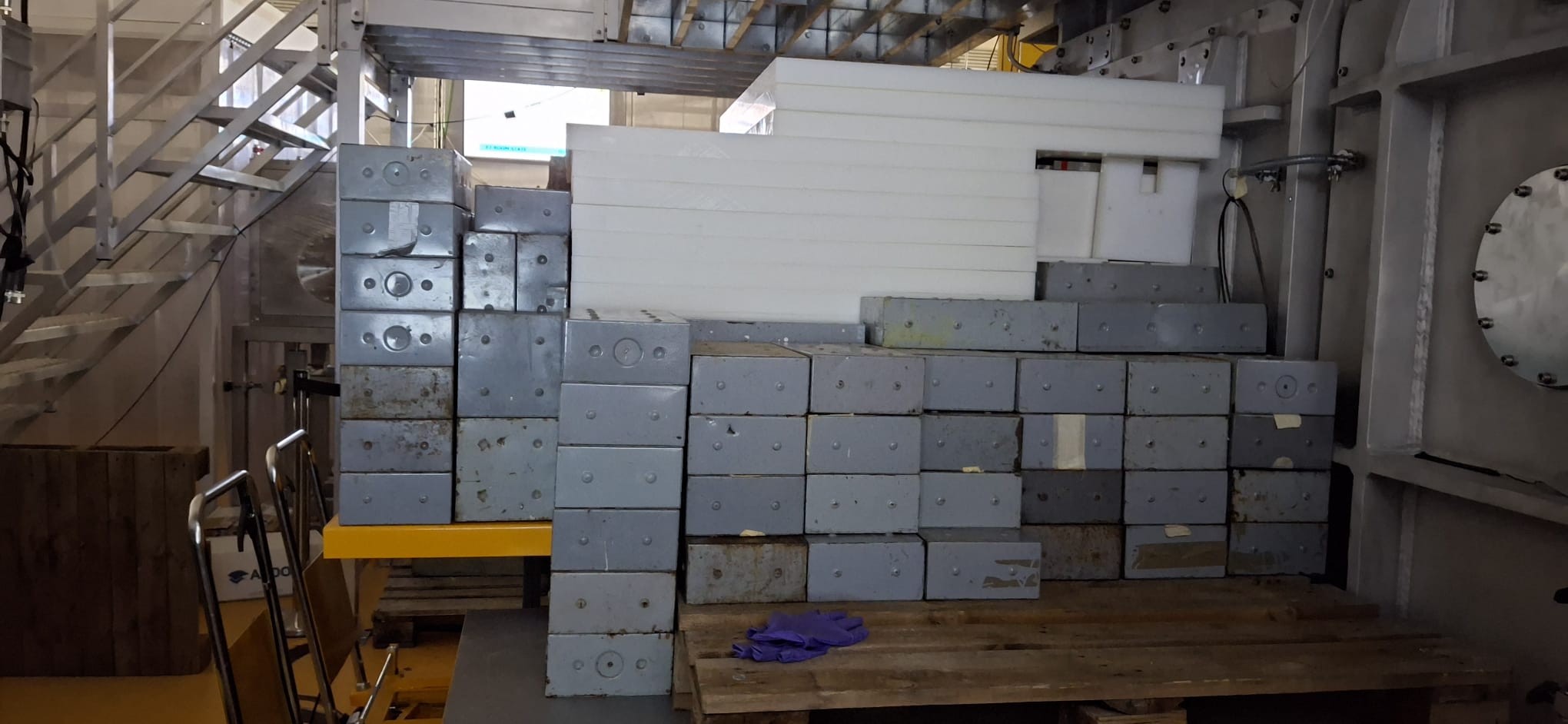}}
\includegraphics[width=0.42\linewidth]{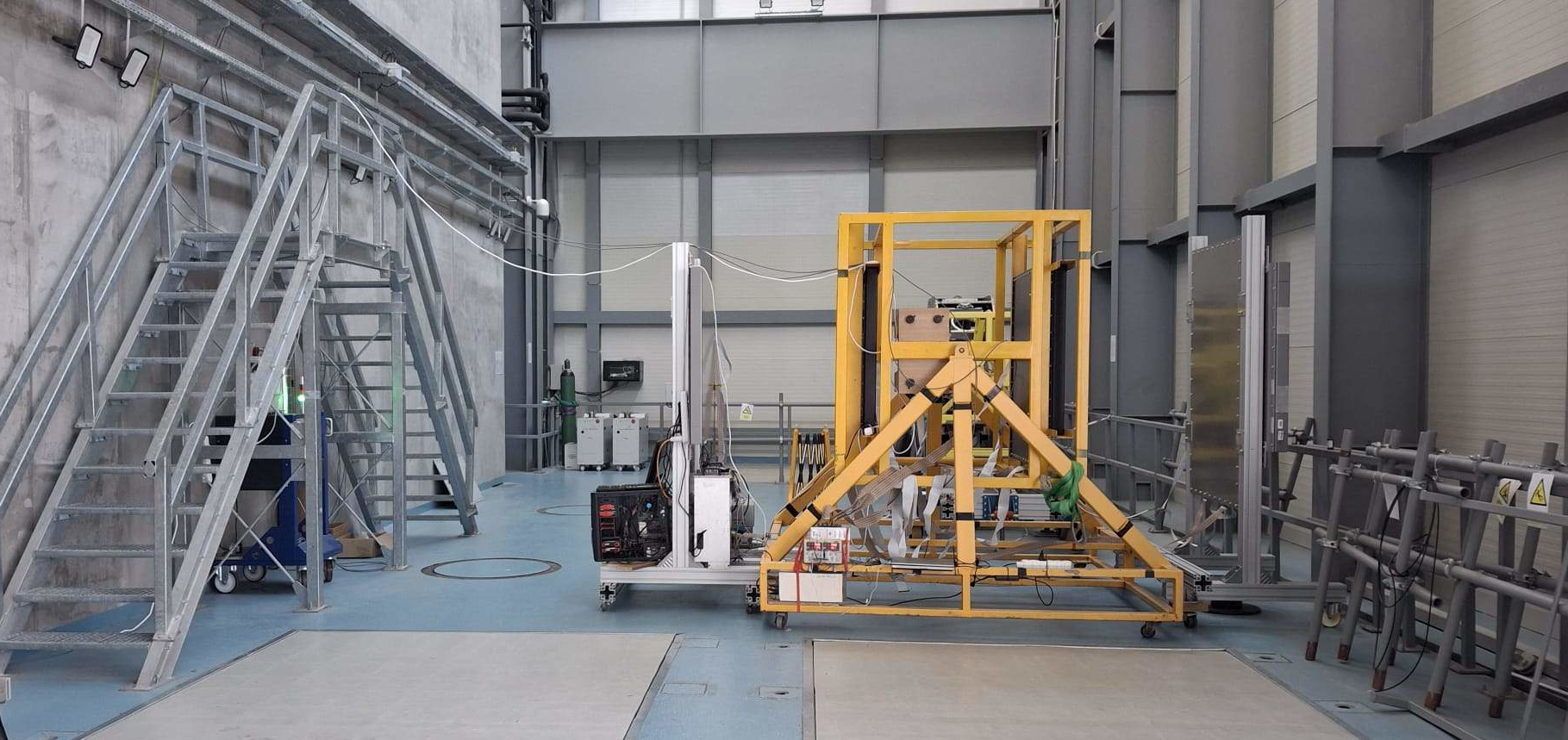}
\includegraphics[width=0.122\linewidth]{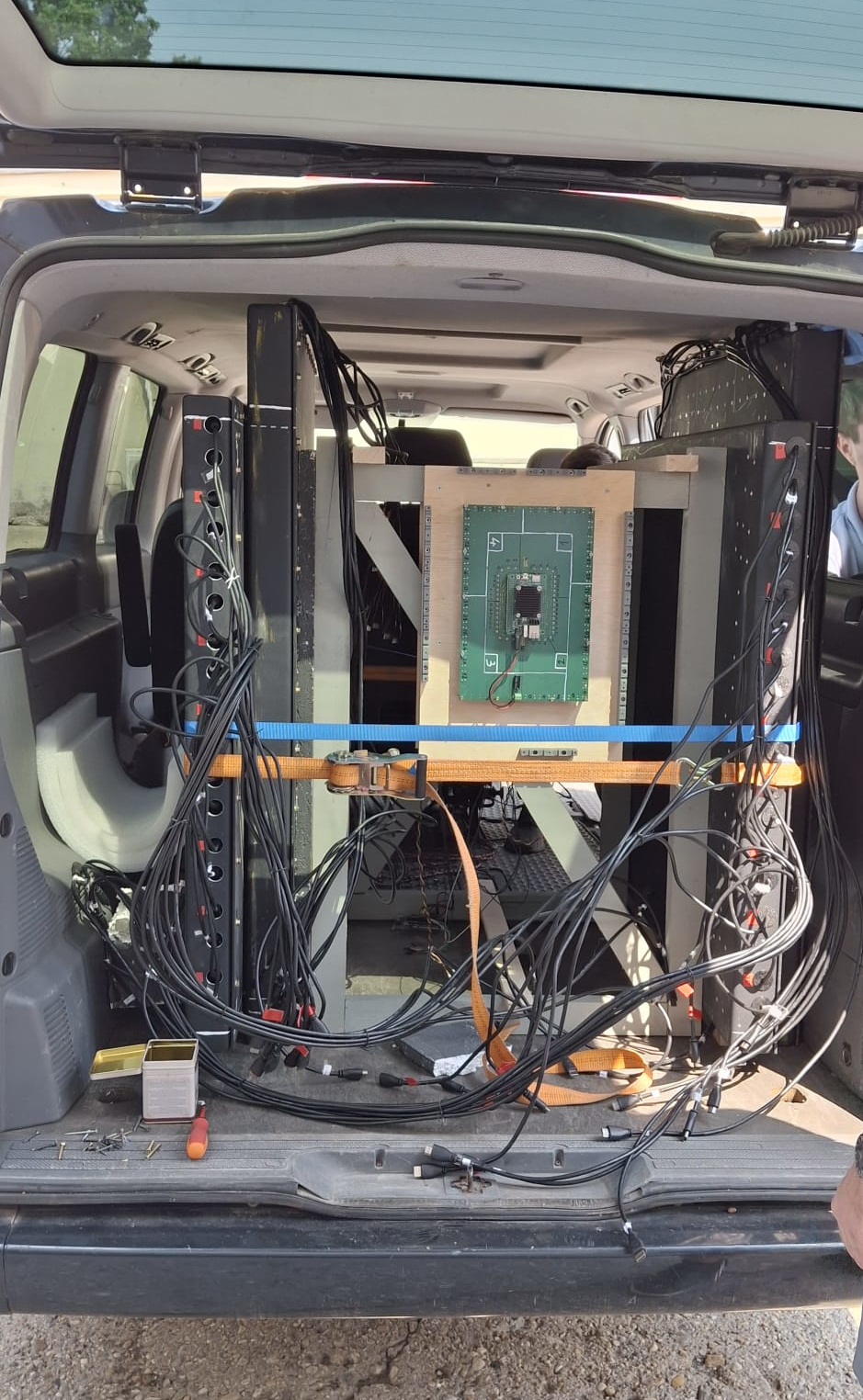}
\caption{{\em Top}: Schematic of the experimental set-up (not to scale), showing the converter target, building walls, and the detectors: T32, SiRO and $\mu$36. The $\mu$36 detector was moved relative to the beam axis, on- and off-axis: at 20 m (red squares, $\pm$1 m, $\pm$2.5 m, $\pm$3.5 m and $\pm$5 m off-axis), at 29 m (blue squares, $\pm$2 m, $\pm$4 m, $\pm$6 m off-axis), and at 42 m (green square) away from the converter. {\em Bottom left:} Partially assembled shielding during the set-up phase. Paraffin bricks (gray boxes) and polyethylene sheets and bricks (white) are placed around the converter lead target. {\em Bottom center:} The SiRO detector (2 sets of detection modules, 1 m apart, mounted on the yellow frame) and T32 modules, placed 1 meter in front of and behind SiRO. The 2-meter concrete wall separating the experimental area from the detectors is located to the left of the image. {\em Bottom right:} The $\mu$36 detector assembled inside a van, with the detection modules oriented perpendicular to the beam axis.}
\label{fig:setup}
\end{figure}

The primary challenge in characterizing muons generated by the interaction of energetic electrons with a solid target is to reliably distinguish them from other secondary particles produced simultaneously, mainly gamma rays and neutrons, which constitute the background signal.
To address this issue and determine clear properties of the muon beam such as angular distribution and energy, an optimized experimental configuration was adopted, incorporating appropriate shielding and a sufficiently large separation between the converter target and the detectors. This arrangement was designed to achieve an optimal balance between suppressing the background to a level that enables unambiguous muon detection, while preserving a muon flux above the detector threshold.
Measurements were performed at the ELI-NP facility in Romania, employing the 10 PW laser arm~\cite{Lureau_Matras_2020,tanaka2020current}, a Ti:Sa system with a central wavelength of 810 nm, that can deliver 230 J on target in 23 fs at a maximum rate of 1 shot/min.
The experimental set-up consists of a 60 mm long gas target for electron-beam generation and a spectrometer to resolve the electron energy spectrum, both placed inside the vacuum chamber, as described in \cite{Diana:2026}. The converter target for the muon generation, a shielding made of various materials, and three detectors were placed outside the vacuum chamber, in controlled areas (Fig. \ref{fig:setup}).
During this experimental campaign, the laser delivered 230 shots. 

The converter target is a 40x40x50 cm$^3$ lead block constructed from 10x10x5 cm$^3$ lead bricks, with a thickness of 40 cm (71.3 $X_0$) along the beam axis. Although muons are mainly generated within the first few centimeters of the target \cite{Calvin:2023,Terzani:2025}, a much thicker target was chosen to attenuate a significant part of the background immediately after muon generation. For such a thick target, electro-production of muon pairs is expected to give significantly lower yields than photo-production or $\pi^{\pm}$/K$^{\pm}$ decays \cite{Titov:2009, Yu:2024}. 

The shielding was configured based on Monte Carlo GEANT4 simulations \cite{Agostinelli:2003,Allison:2016}, designed to slow down and thermalize neutrons and attenuate gamma rays near the converter. The first layer of the shielding was made of stacks of polyethylene sheets and bricks with a total thickness of 100 cm along the beam axis and 30 cm above and around the converter target. The second layer consists of paraffin bricks encapsulating the entire polyethylene structure, with a thickness of 60 cm along the beam axis and 40 cm to the sides and above. 
An additional lead dumper and the 2-meter reinforced concrete wall of the experimental area completed the shielding.

Three different scintillator-based detectors developed for muography applications, namely the T32, SiRO and $\mu$36, have been employed to measure the muon beam.
The T32 and SiRO \cite{siro1:2024} detectors were placed outside the experimental area, 12 m from the converter (behind the 2-meter concrete wall), centered on the beam axis, and kept stationary throughout the campaign to measure muon flux and time of flight. The T32 detector is made of 2 pairs of detection layers placed 3.1 m apart and centered on the SiRO detector (Fig. \ref{fig:setup}, bottom-center photo).

The third detector, the $\mu$36 muon telescope, is specifically designed to reconstruct the trajectory of atmospheric muons \cite{mu36:2025,time_SiPM:2023}. It consists of four detection layers, each made of 36 scintillator bars. For this experiment, the detector was mounted inside a vehicle, with the detection layers oriented vertically and facing perpendicularly to the muon beam axis. The vehicle was located behind a 7-centimeter metal-concrete wall in a controlled area outside the experimental building, and moved along the beam axis and sideways to sample the beam profile (see Fig. \ref{fig:setup}). 

\section{Monte Carlo simulations}\label{sec:MC}

To fully understand and describe the phenomenology associated with the particles resulting from the interaction of high-energy electrons with a solid target, GEANT4 simulations were performed using version 11.2.2 and the FTFP\_BERT physics list. To achieve a realistic simulation of all particle fluxes, muons and background, the GammaToMuons process (Bethe-Heitler) was activated with the GammaToMuonsFactors kept at 1, and the entire experimental set-up was reconstructed (shielding, detectors, magnet, building elements).

Several sets of simulations have been run to address key factors: 1) determine the muon yields and the contribution of the generation process to the muon fluxes that reach the detectors, 2) determine the energy spectrum and spatial distribution of background particles, 3) determine the energy deposited in the detectors by the incident particles to calculate detector response.

\begin{figure}[tb!]
\centering
\includegraphics[width=0.4\linewidth]{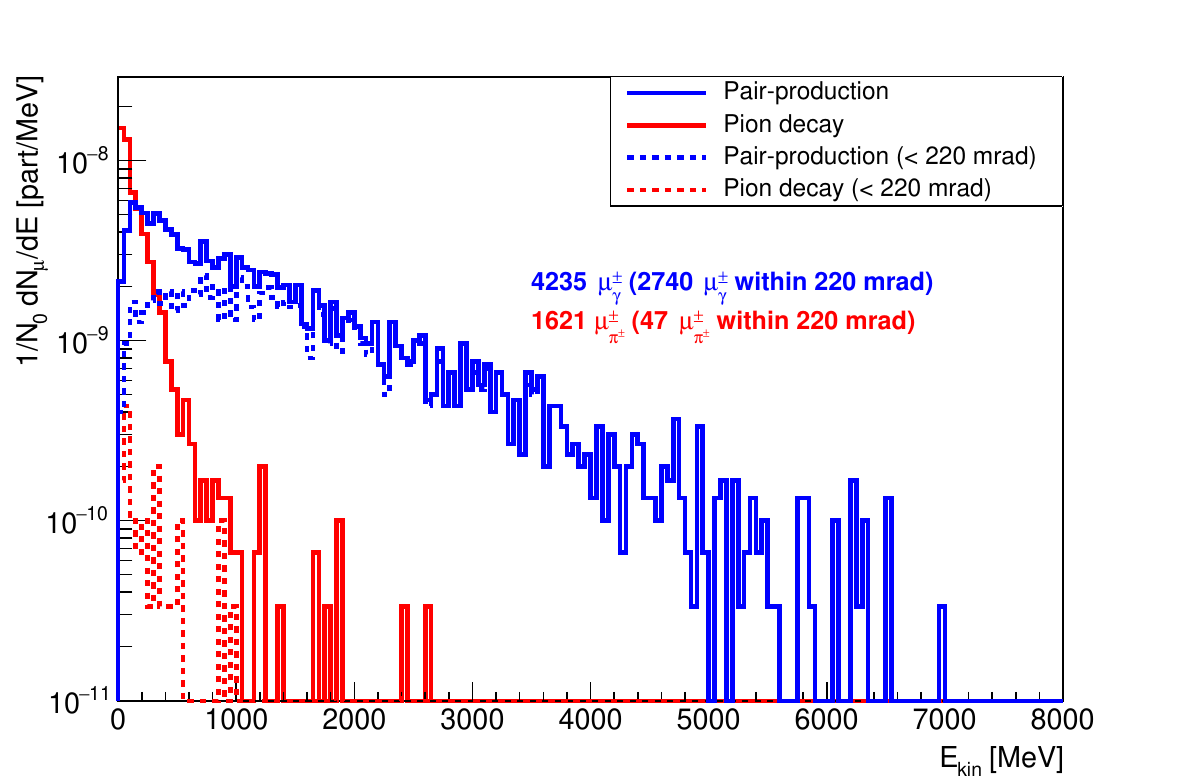}
\caption{Simulated energy spectra of muons generated by pion decays (red) and pair-production (blue) by an 8$\pm$1 GeV electron beam ({\em i.e.} 6$\times$10$^8$ primary electrons) incident on the lead target, for all muons (solid lines) and only muons with forward directions contained within a $\sim$13$^\circ$ full-cone angle centered around the beam axis, an angle which corresponds to the aperture of the converter target used in the experiment (dashed lines).}
\label{fig:mu_types}
\end{figure}

First, considering a primary electron beam (6$\times$10$^8$ primary electrons) with a narrow energy band of 8$\pm$1 GeV, which is the maximum electron energy obtained in this work, the muon yield was accurately evaluated by simulations to underline the difference in flux and angular distribution between the muons generated via the Bethe-Heitler process (pair-production) and those resulting from pion/Kaon decays. As illustrated in Fig. \ref{fig:mu_types}, Bethe-Heitler muons dominate the yield. They are significantly more energetic than those from pion decays, and are highly collimated, with a substantial fraction falling within a solid angle of $\sim$40 msr around the beam axis (\textit{i.e.} $\sim$13$^\circ$ full-cone angle, the aperture of the converter target). Therefore, according to simulations, muons generated via pair production are essentially the only ones able to exit the experimental area and reach the detectors, along with neutrons and secondary gamma photons.
For the second set of simulations, measured electron spectra were used as input \cite{Diana:2026} in order to accurately reproduce the particle species, energies, and spatial distributions of the particles generated in a pulse. However, simulating each shot individually is a difficult computational process in terms of both processing time and storage space. To simplify the procedure, we developed a three-stage algorithm based on energy binning and statistical scaling that provides the accurate particle distributions and energy spectra generated in each pulse without simulating individual pulses.
In the first stage, the electron spectrum from 2 to 9 GeV was discretized into 0.5 GeV energy bins. For each bin, $i$, 1.5 $\times$ 10$^8$ primary electrons ($N_{sim, i}$), with energies uniformly distributed within the bin, were independently simulated. 
The second stage consists of applying the same binning to the experimentally measured electron spectra, that is, using the same energy range and bin width. For each laser shot and energy bin, the number of primary electrons $N_{exp, i}$ was thus extracted from the measured spectrum.
In the third stage of the procedure, each simulated bin was rescaled with respect to the corresponding bin of the measured spectrum to reconstruct the entire electron energy spectrum, and consequently the energy spectrum of the secondary particles.
Thus, any experimental spectrum may be reconstructed as a linear combination of all bin spectra weighted with appropriate coefficients. 
This method allows for a precise characterization of secondary particle production under conditions that replicate experimental reality while remaining independent of the spectral characteristics of each individual shot, and removing the need to perform a simulation for each specific electron spectrum.
With this method, all shots from the experimental campaign were reconstructed. 

\begin{figure}[b!]
\centering
\includegraphics[width=0.7\linewidth]{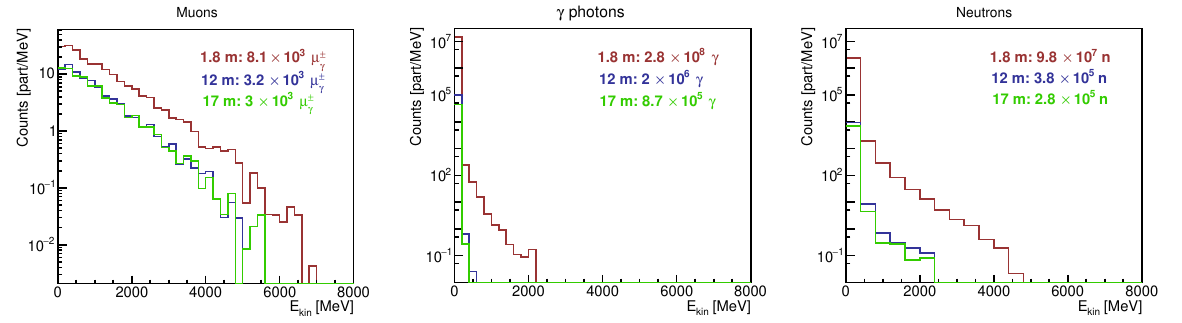}
\caption{Simulated energy spectra of muons (left), gamma rays (center), and neutrons (right). The simulation was initialized with an electron spectrum obtained by averaging more than 30 laser shots.
The spectra are retrieved at three different positions: at the exit of the converter target and polyethylene-paraffin shielding, 1.8 m from the converter (red lines); at 12 m from the converter, where the first two detectors are placed (blue lines); and 17 m away from the converter, outside the experimental building (green lines). The numbers on the plots represent the total number of particles for each species and per laser shot.}
\label{fig:background_spec}
\end{figure}

According to the simulations, most of the background comes from neutrons and secondary gamma photons produced in the 2-meter concrete wall. Fig. \ref{fig:background_spec} shows the simulated energy spectra of muons, gamma photons, and neutrons at three different locations from the converter: at 1.8 m from the converter (at the exit of the shield surrounding the converter target), at 12 m (the position of the first detector), and at 17 m (after the thin metal concrete wall). As the primary particle source, the simulation used an electron spectrum obtained by averaging more than 30 laser shots.
A final series of simulations was performed to assess the energy deposited in the detector at various positions.

\section{Muon beam profile measurement}\label{profile}

\begin{figure}[tb!]
	\centering
	\includegraphics[width=0.65\linewidth]{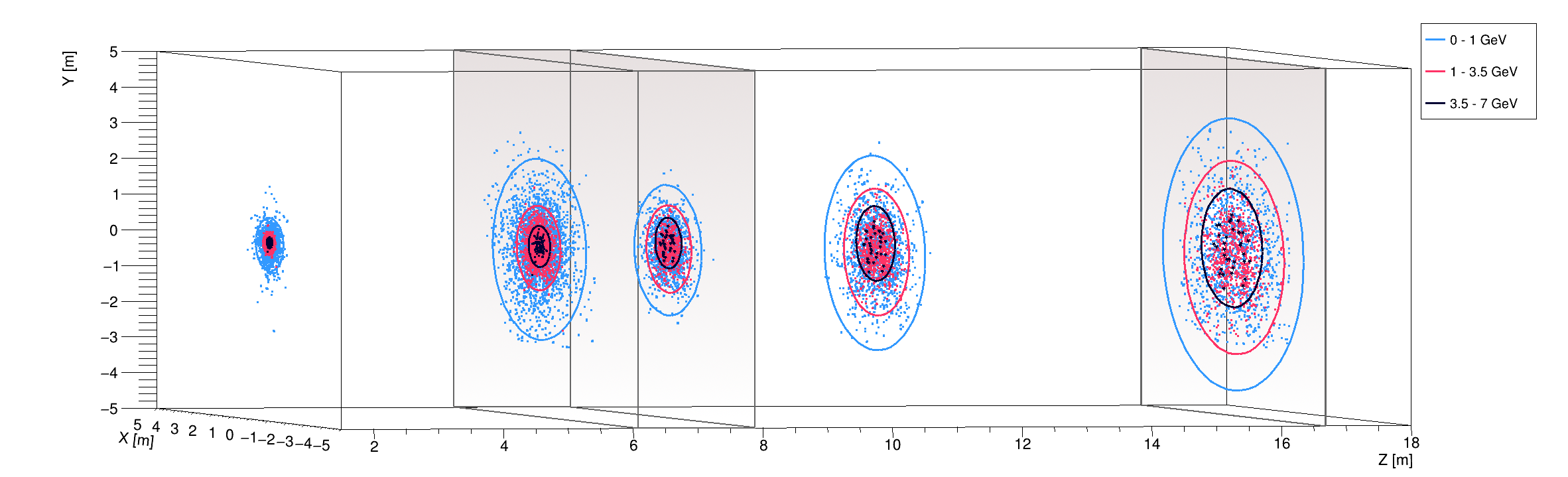}
	\includegraphics[width=0.65\linewidth]{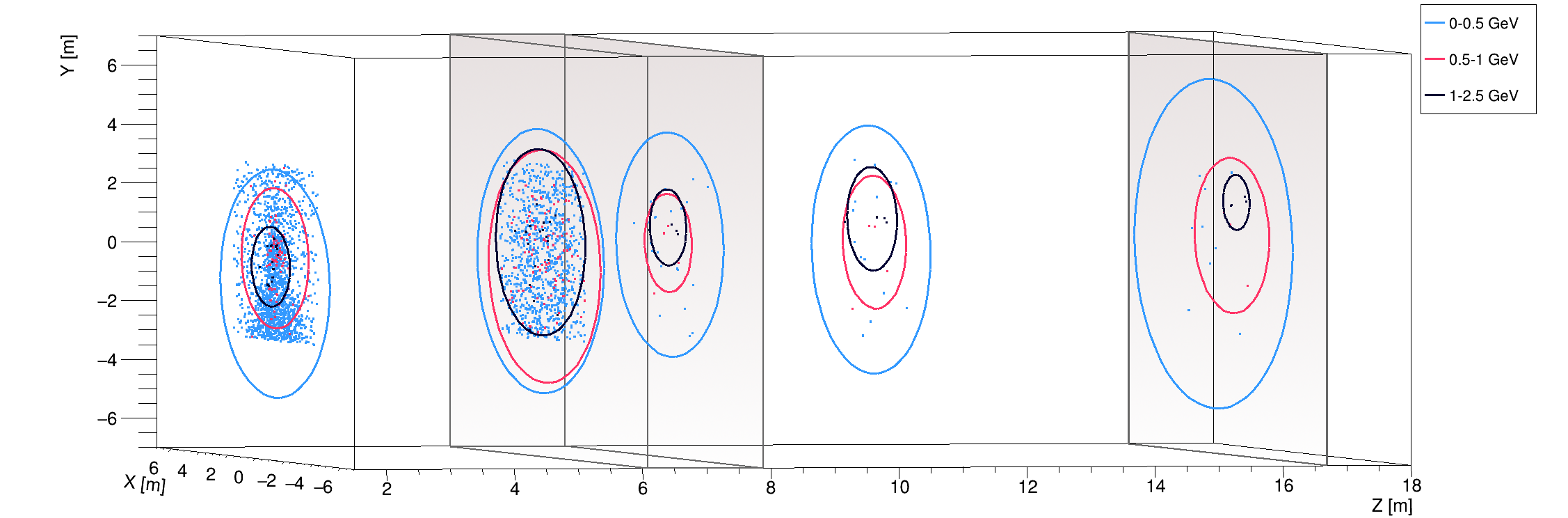}
	\caption{Simulated spatial distribution of gamma-generated muons (top) and muons resulting from pion decays (bottom) at various distances from the interaction chamber according to their energies at the respective distance. The circles show the area into which 98.9\% of particles are contained ($3\sigma$ threshold of the corresponding Rayleigh distribution), displayed as circles in the corresponding colors. Real measured electron pulses have been used as input for the GEANT4 simulation. The two gray blocks represent the existing walls within the experimental setup.} 
	\label{fig:muon_spread}
\end{figure}

As muons propagate from the target to the detectors, they slow down and scatter by an angle that depends on the target material and the muon energy. 
In the current experimental set-up, consisting of several types and sizes of shielding materials (\textit{e.g.}, polyethylene, paraffin, concrete walls), muons will encounter multiple scattering regions along their propagation from the target to the detectors. 
This increases the divergence of the muon beam from its initial divergence at the converter target. According to simulations, the muon beam is expected to spread over a few meters at the position of the $\mu$36 detector, as shown in Fig. \ref{fig:muon_spread}.

\begin{figure}[htb]
\centering
\includegraphics[width=0.7\linewidth]{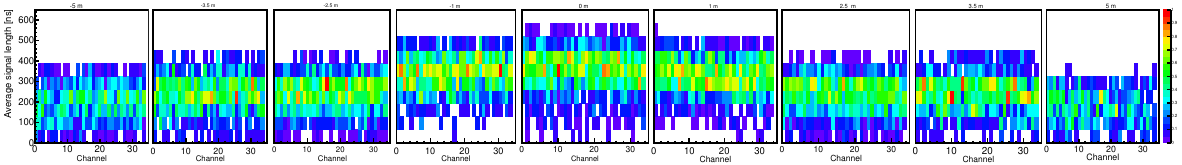}
\includegraphics[width=0.7\linewidth]{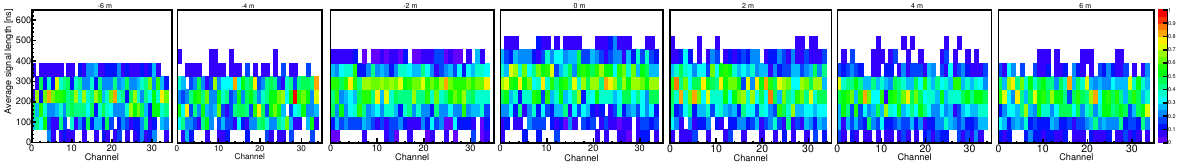}
\includegraphics[width=0.5\linewidth]{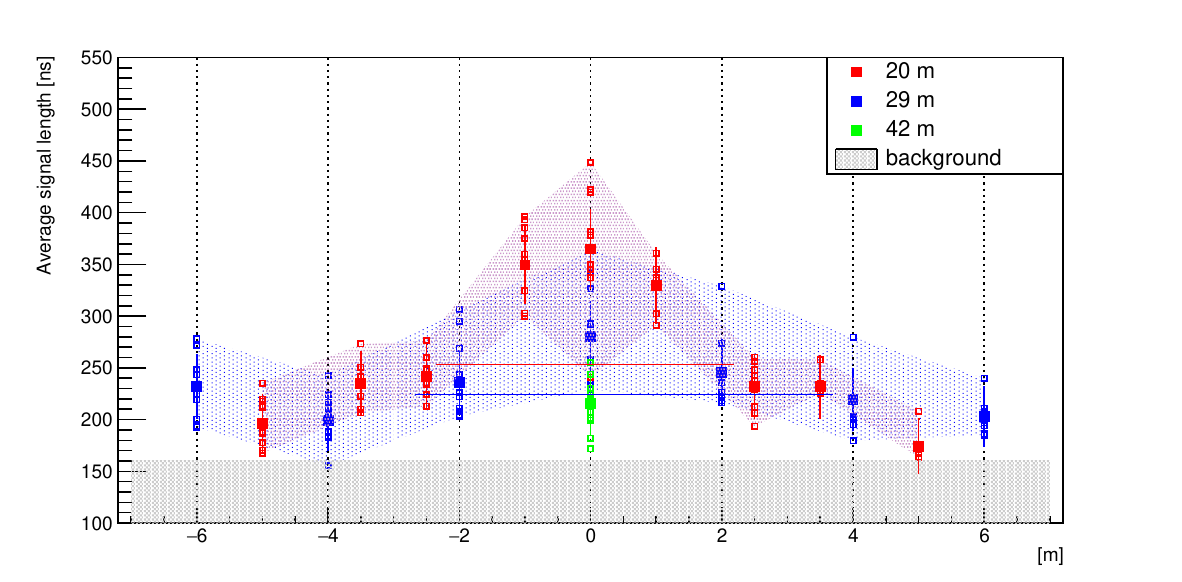}
\includegraphics[width=0.5\linewidth]{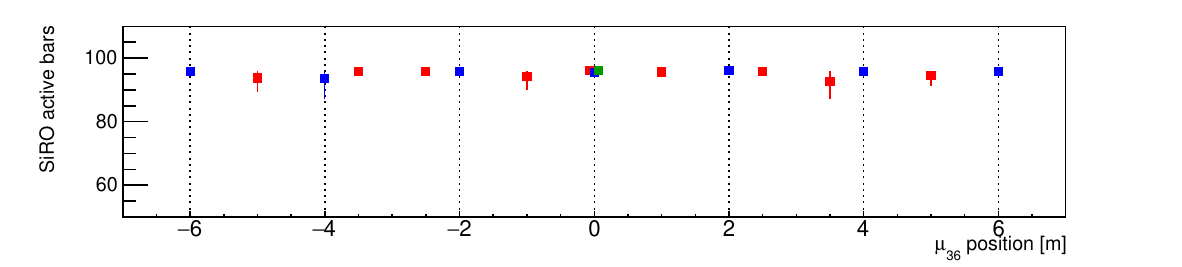}
\caption{{\em Top:} Measured signal length, in nanoseconds, per bar, averaged for the 10 pulses recorded for each position, by the $\mu$36 detector placed at 20 m (red points), 29 m (blue points), and 42 m (green point) away from the converter, left and right of the beam axis in the positions displayed in figure \ref{fig:setup}. {\em Center:} Signal length, in nanoseconds, averaged for all the bars of the $\mu$36 detector, at the measuring positions. Empty squares indicate individual pulses (10 for each position) while filled squares indicate the average, with error bars indicating the standard deviation induced by he variability of the electron source. The background is given by the measured signal length of individual cosmic muons, with a maximum at 160 ns. {\em Bottom:} Number of active bars in the SiRO detector (count rate) during the measurements performed with the $\mu$36 detector. 
The {\em x}-axis corresponds to the position of the $\mu$36 detector, while the SiRO detector was stationary on the beam axis 12 m away from the converter.}
\label{fig:signal_length_all}
\end{figure}

\begin{figure}[htb]
\centering

\includegraphics[width=0.6\linewidth]{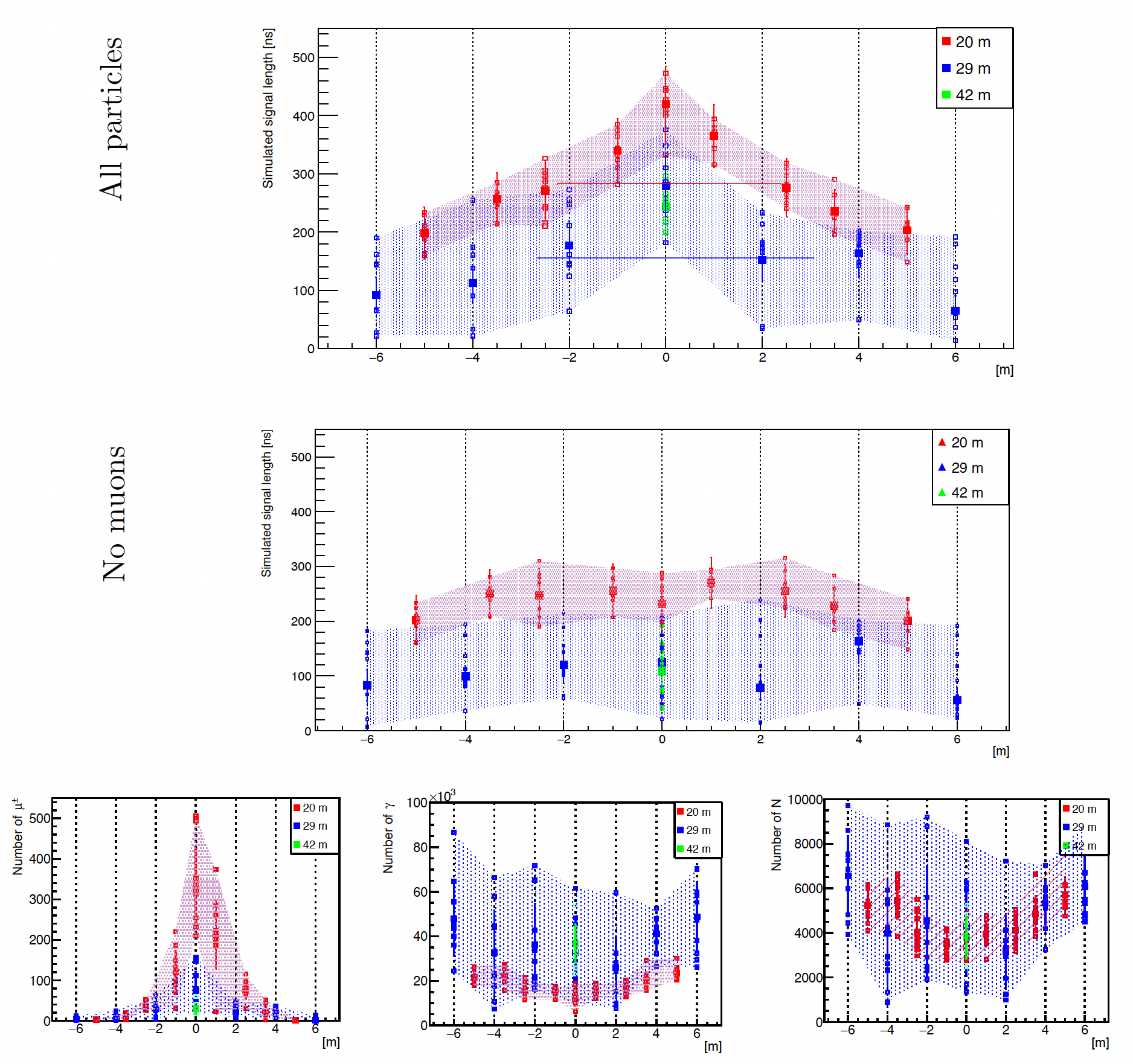}

\caption{{\em Top:} Estimated signal length obtained from the simulated energy deposited by all particles in the $\mu$36 bars, and at the same positions of the measurements (20, 29 and 42 m from the converter, and up to 6 m left and right). The signal is corrected for the time-over-threshold effect in the electronics and for the 10 pulses recorded by the detector at each position. {\em Center: } Estimated signal length obtained from the simulated energy deposited in the $\mu$36 bars, corrected for the time-over-threshold, for the 10 pulses recorded by the detector at each position, excluding the muons. 
{\em Bottom:} Simulated number of particles (muons, gamma photons, neutrons), per pulse, incident on an area of 1x1 m$^2$ (the detector's area). Errors represent the standard deviation.}
\label{fig:signal_length_all_sim}
\end{figure}

To sample the muon beam profile, the $\mu$36 detector was placed at 20 m, 29 m, or 42 m from the converter, and moved left and right of the beam axis up to 6 m, in the positions illustrated in Fig. \ref{fig:setup}.
Figure \ref{fig:signal_length_all} shows the measured signal length for each position (10 pulses per position). Top panels display signal length for individual bars across the surface of the detector, averaged over the 10 pulses recorded at each position. 
The central plot shows the signal length over the surface of the detector, for individual pulses (empty squares) and averaged over the 10 pulses recorded at each position (filled squares with standard deviation).
The bottom plot shows the count rate of the SiRO detector that was kept stationary, at 12 m away from the converter, serving as a beam monitor.
It can be seen that there is a peaked distribution in the $\mu$36 signal, centered on the beam axis, with a FWHM of 4.51 m for the points at 20 m from the converter and 6.35 m for the measurements performed at the 29 m distance, which is not observed in the stationary detector.
To understand the results, we have performed GEANT4 simulations of the complete experimental set-up, using as starting point the measured electron spectra of the corresponding pulses recorded by the $\mu$36 detector. 
Considering the standard physics behind the simulation processes (fully implemented in GEANT4 up to 100 TeV for muons and their generators \cite{Allison:2016,Agostinelli:2003}) and also that simulations have been previously benchmarked against similar measurements and found to be in good agreement with recorded data \cite{Battaglieri:2019}, we use GEANT4 simulations to compare and validate our measurements. Since the measured signal length is a function of the energy deposited in the scintillator by the incoming particles, we have extracted from the simulations the deposited energy. To include detector effects, the deposited energy was subsequently corrected for the light collection factors and the time-over-threshold (ToT) effect caused by the detector's electronics, to obtain the estimated signal length.
Figure \ref{fig:signal_length_all_sim}, top panel, displays the simulated signal length with a FWHM of 4.43 m for the pulses observed at 20 m, and 6.27 m for 29 m.

\begin{figure}[ht]
\centering
\includegraphics[width=0.6\linewidth]{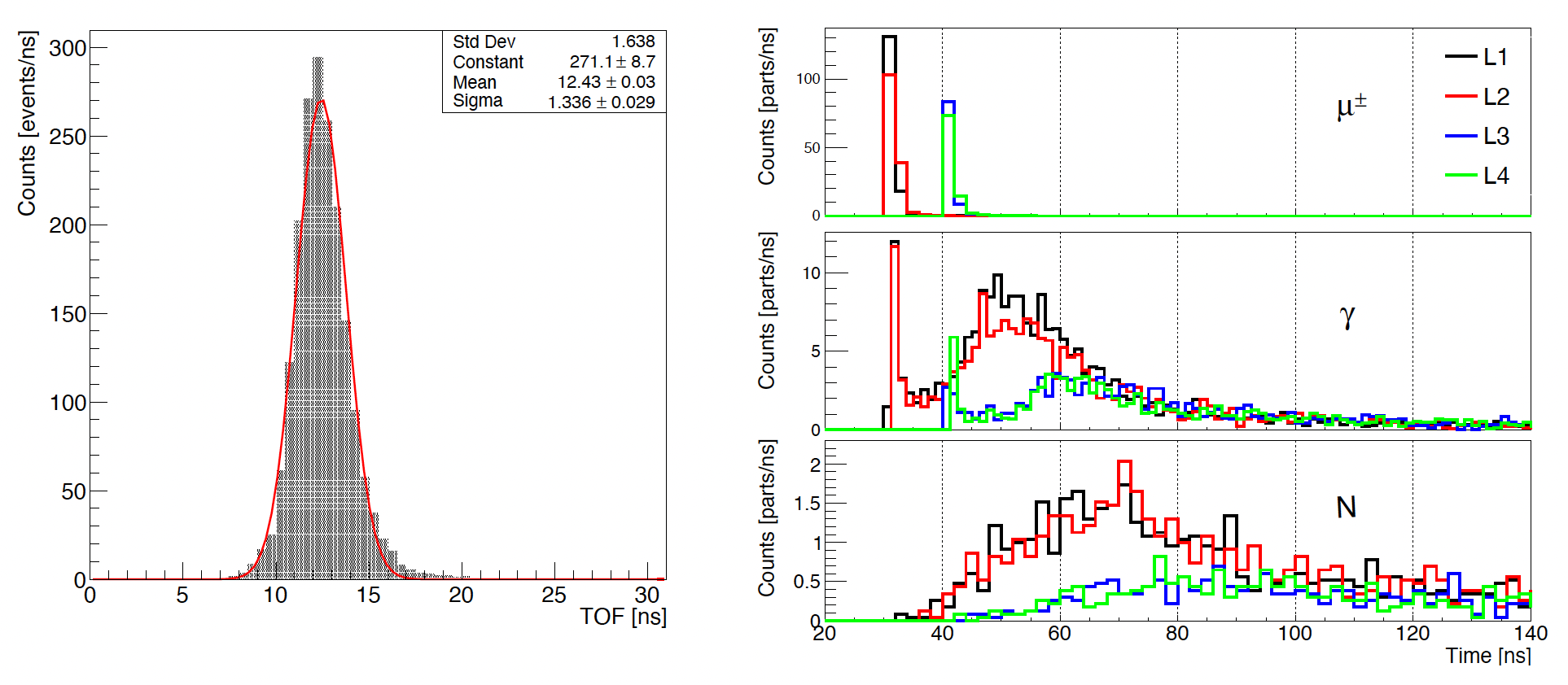}

\caption{{\em Left:} Distribution of time-of-flight between the first and last detection layer of the T32 detector placed at 12 m from the interaction chamber, considering only the first hit in each bar. {\em Right:} Simulated arrival time at the four plates of the T32 detector for muons, gamma photons and neutrons. Colors represent the four detection plates. }
\label{fig:tof}
\end{figure}

In order to evaluate the contribution of each particle type to the overall signal, the number of muons, gamma photons, and neutrons incident on the $\mu$36 sensitive area (1$\times$1 m$^2$) were simulated (see Figure \ref{fig:signal_length_all_sim}, bottom plots). It can be observed that a significant number of muons is present for the on-axis measurement and decreases at the off-axis positions (peaked distribution), while the number of neutrons/gamma photons is mostly constant, with a slight increase at large distances due to the asymmetrical shielding (deeper on the forward direction and shallower on the sides of the converter target).
The simulations also predict more background particles than muons, which may lead us to expect that the muons' contribution to the overall signal would be insignificant. However, the detector response to the different types of particles varies. 
Muons, in the minimum-ionizing-particle (MIP) regime, which is the case of this experiment, deposit in the polystyrene-based scintillator ($\rho$=1.06 g/cm$^3$) 1.936 MeV cm$^2$/g \cite{Groom:2001}, while simulations show that the keV-MeV gamma photons and sub-GeV neutrons incident on the detector deposit much less energy. At 20 m from the converter, and after the shielding, simulations show that 86.91\% of the energy deposited in the scintillator comes from muons, while only 13.09\% is attributed to other particles.
We conclude that the observed beam profile is consistent with the simulations and is dominated by muons, while other particles' contributions are much smaller.
Also stationary, the T32 detector measured the time of flight between the first and last plate, placed 3.1 m apart from one another. Measurements presented in Fig. \ref{fig:tof} give a ToF of 12.43 ns with a $\sigma$ of 1.34, while the corresponding simulations give a value of 10.02 ns. 

\newpage

\section{Muon energy estimate}\label{energy}

\begin{figure}[ht]
\centering
\includegraphics[width=0.25\linewidth]{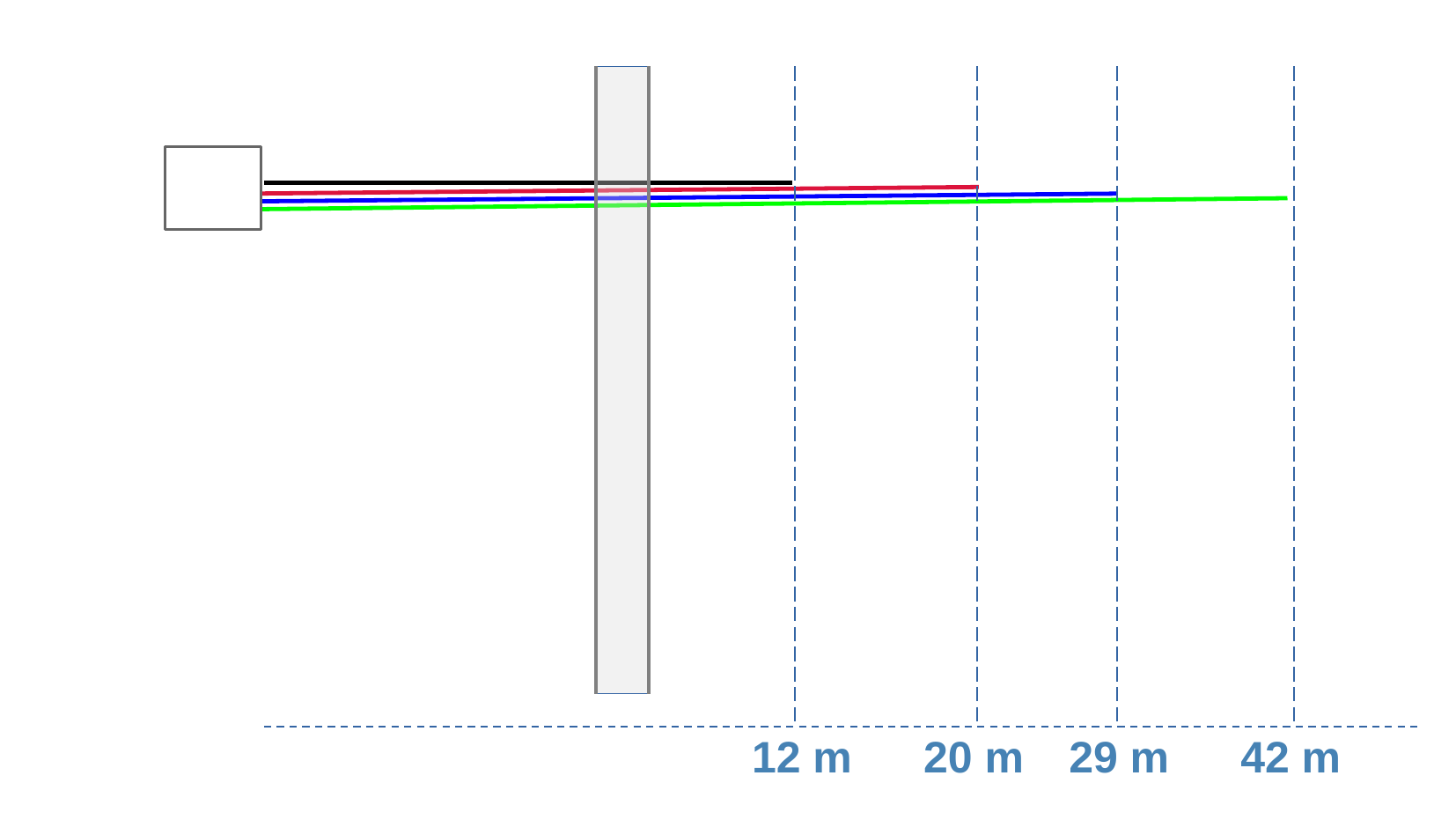}
\includegraphics[width=0.25\linewidth]{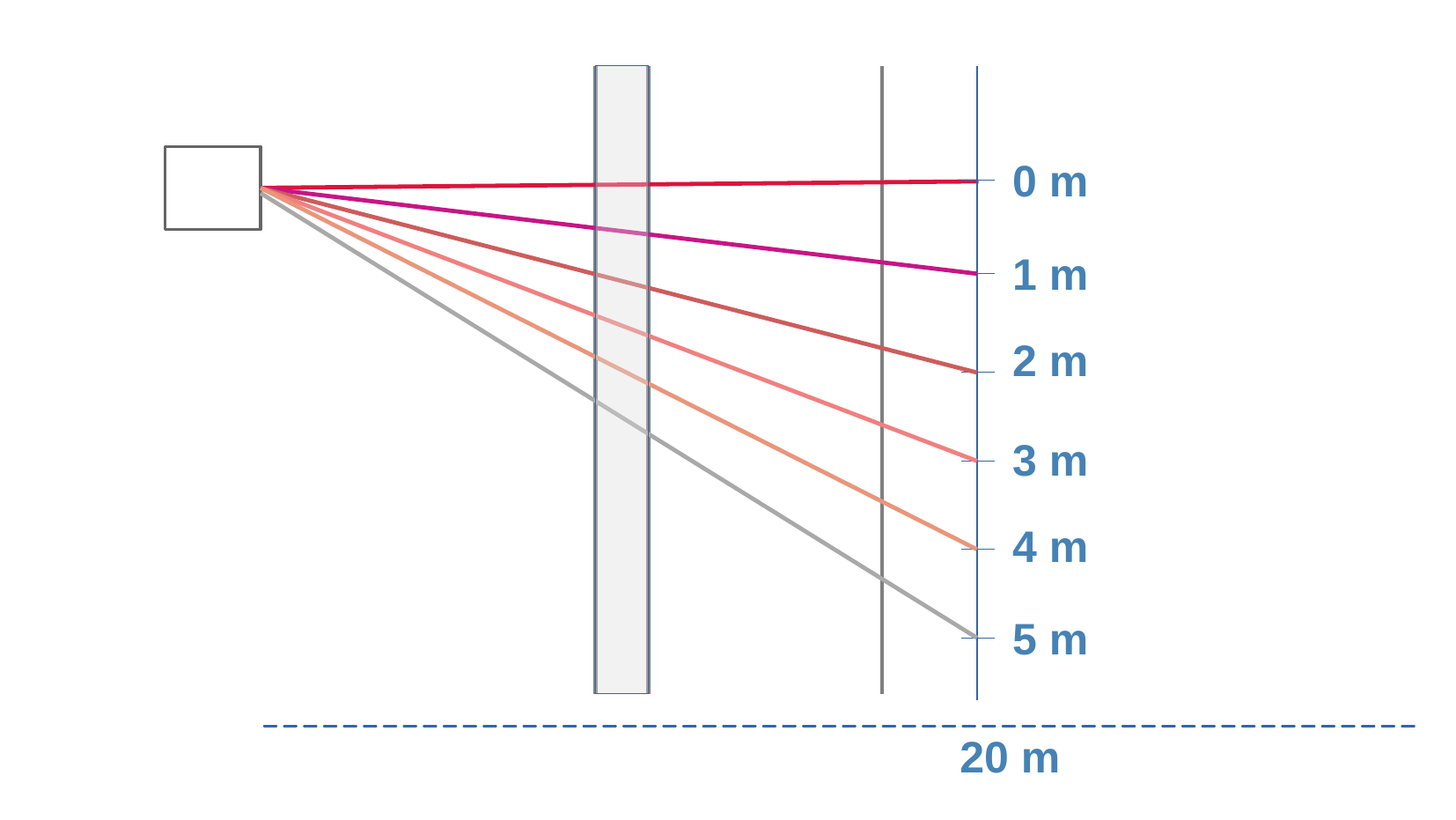}
\includegraphics[width=0.25\linewidth]{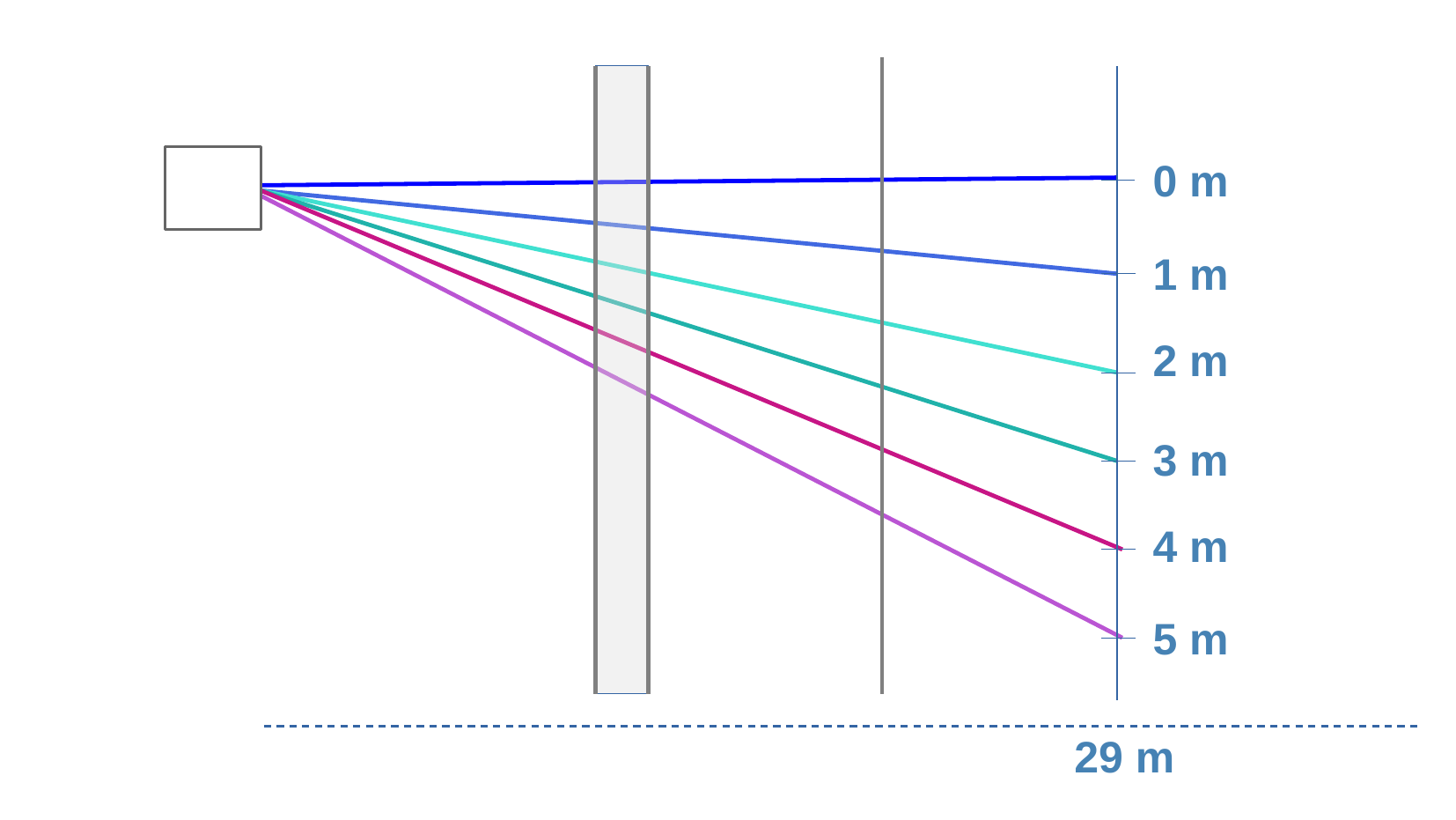}
\\
\includegraphics[width=0.25\linewidth]{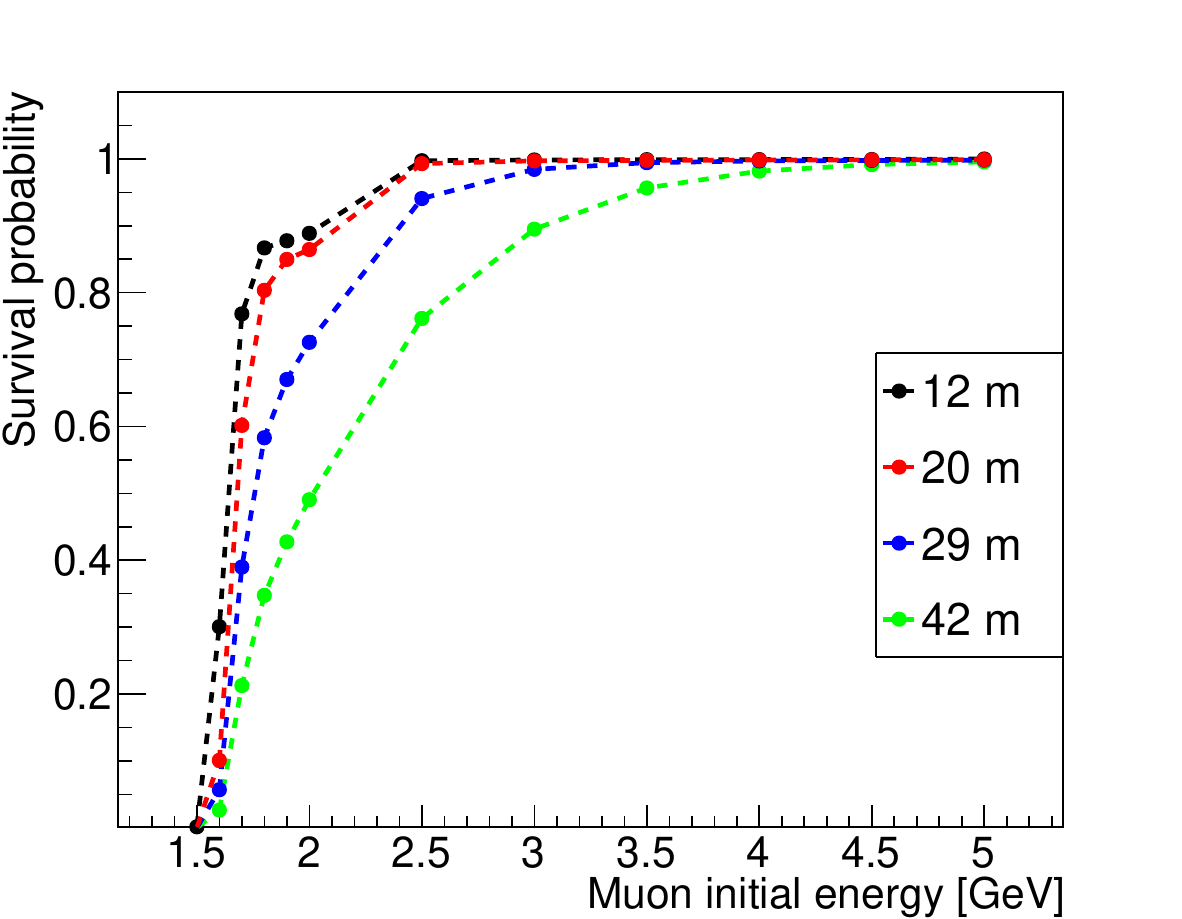}
\includegraphics[width=0.25\linewidth]{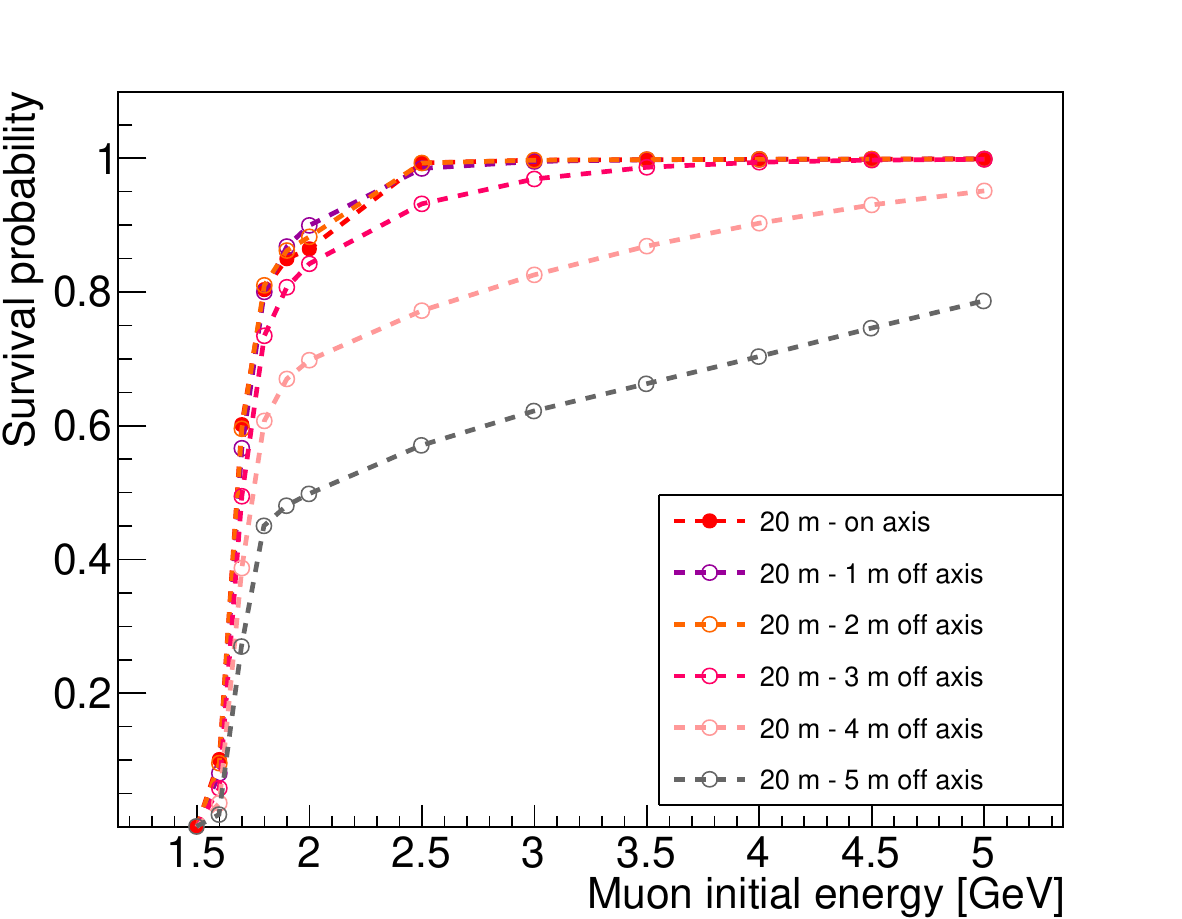}
\includegraphics[width=0.25\linewidth]{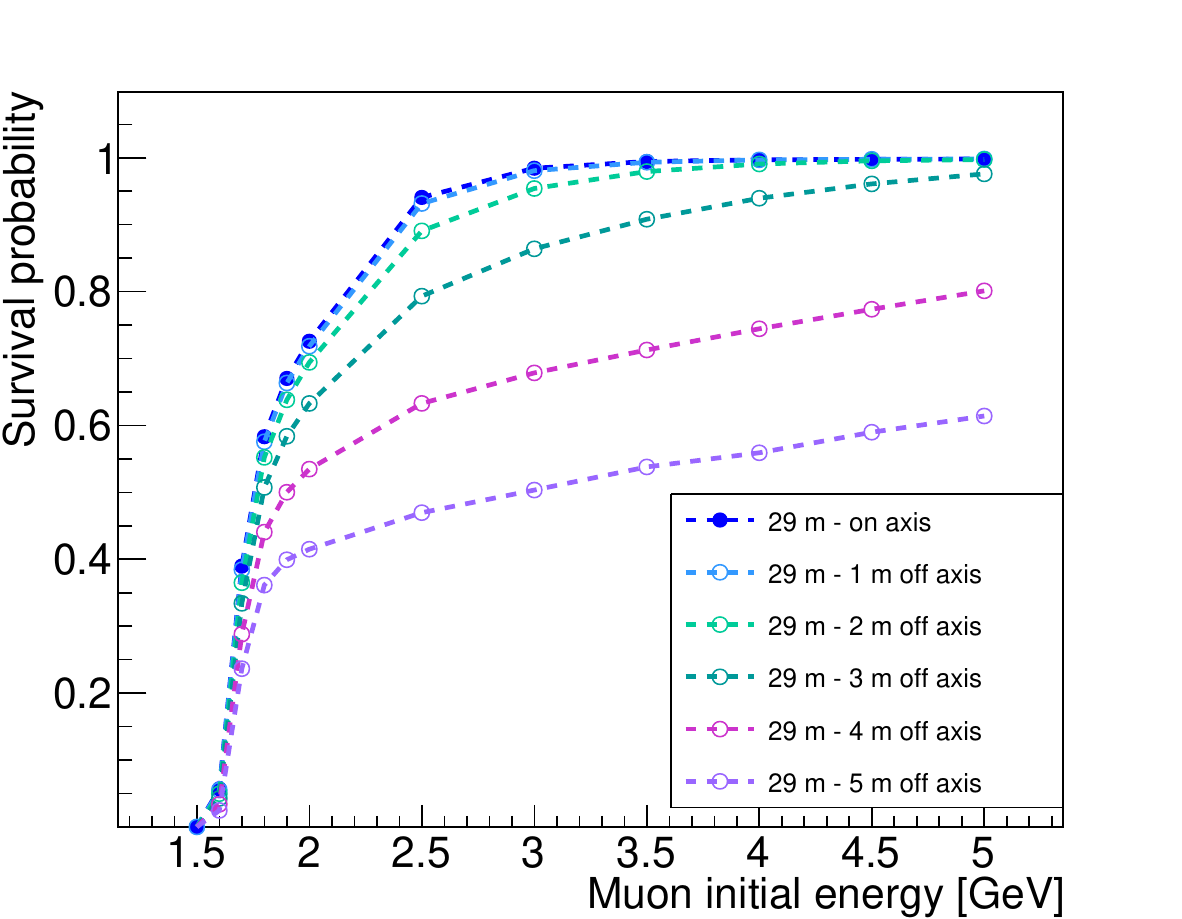}
\caption{(Top) Schematics show the top view of the paths muons travel through the experimental setup to each measuring position. The gray area marks the 2-meter reinforced-concrete wall. (Bottom) Simulated survival probability of muons as a function of their initial energy, for muons traveling along the beam axis (left), muons that reach as far as 20 m from the converter and up to 5 m off-axis (center), and muons that reach as far as 29 m from the converter and up to 5 m off-axis (right).}
\label{fig:min_energy}
\end{figure}

To investigate the primary energy needed for a muon to reach the measuring locations, we have injected monochromatic muons of 1.5, 1.6, 1.7, 1.8, 1.9, 2.0, 2.5, 3.0, 3.5, 4.0, 4.5, and 5.0 GeVs into a GEANT4 simulation of the experimental set-up.
Figure \ref{fig:min_energy} shows the survival probability of the monochromatic muons at each distance where laser-induced particle bunches have been experimentally measured, on-axis and off-axis up to 5 m.
The simulation shows that muons with energy below 1.5 GeV cannot reach any detectors, while muons with energy above 3 GeV have a probability close to 1 of reaching all detectors.

\section{Muon imaging}\label{transmission}

\begin{figure}[ht]
\centering
\includegraphics[width=0.38\linewidth]{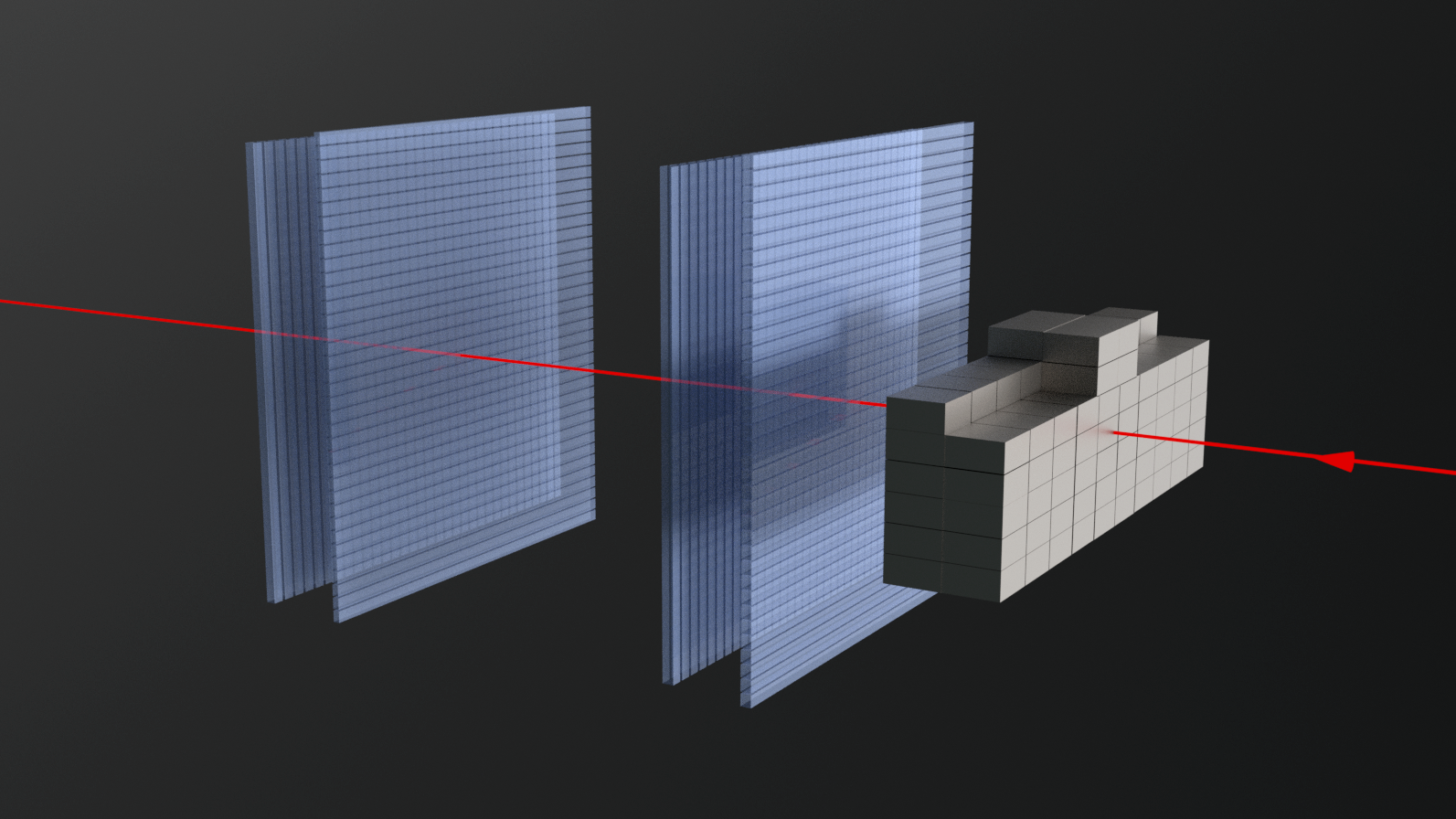} 
\includegraphics[width=0.3\linewidth]{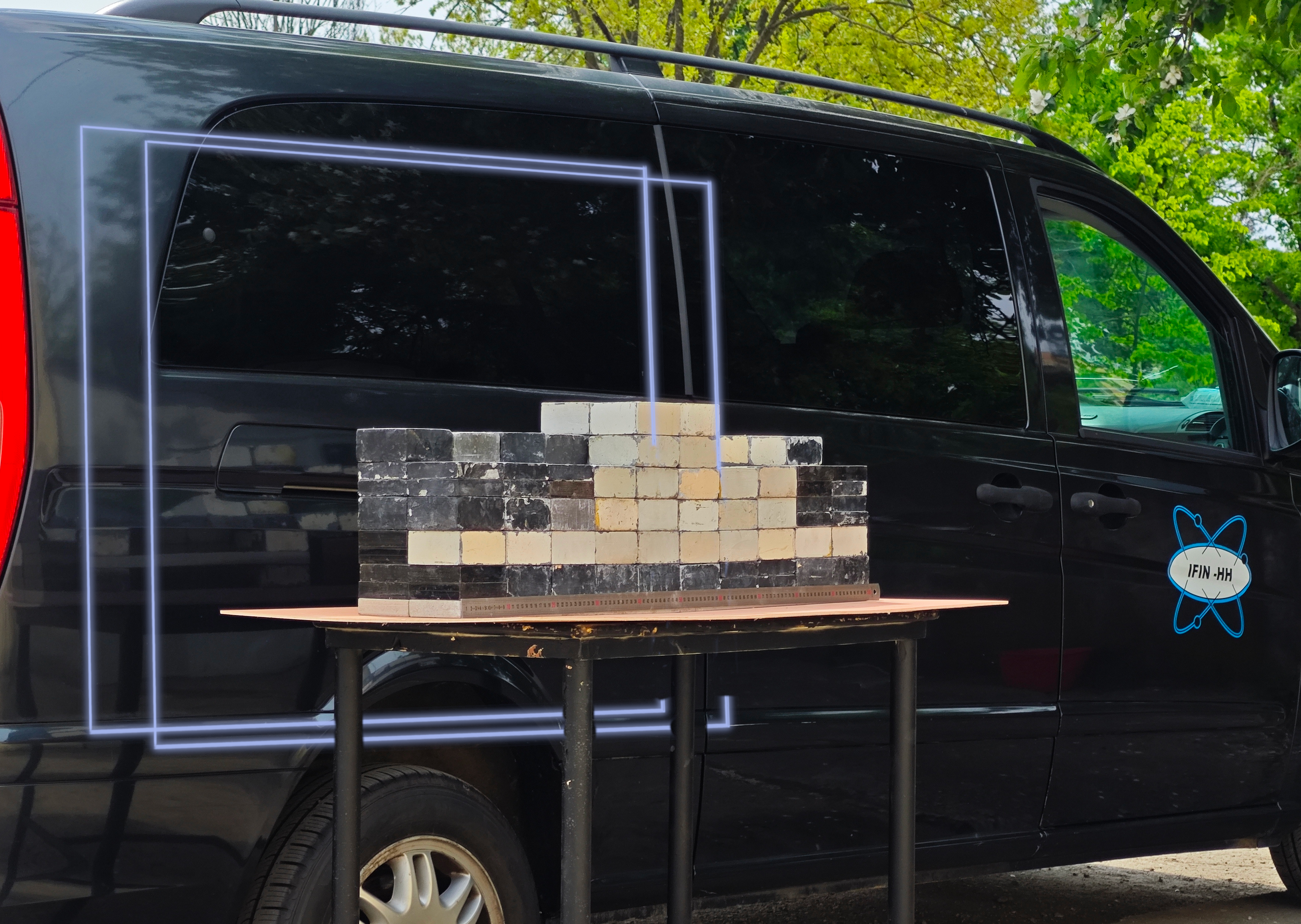}

\includegraphics[width=0.18\linewidth]{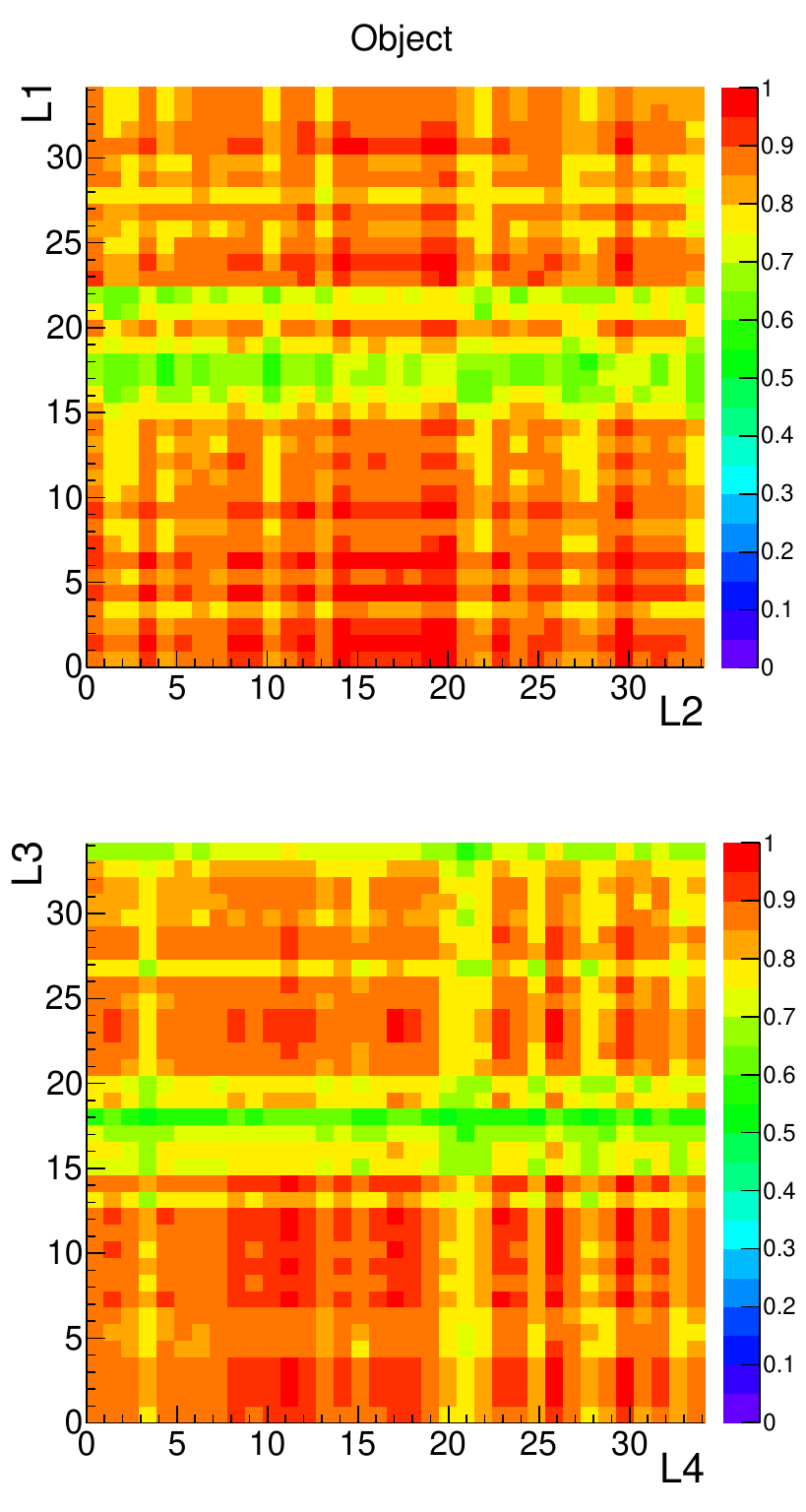}
\includegraphics[width=0.18\linewidth]{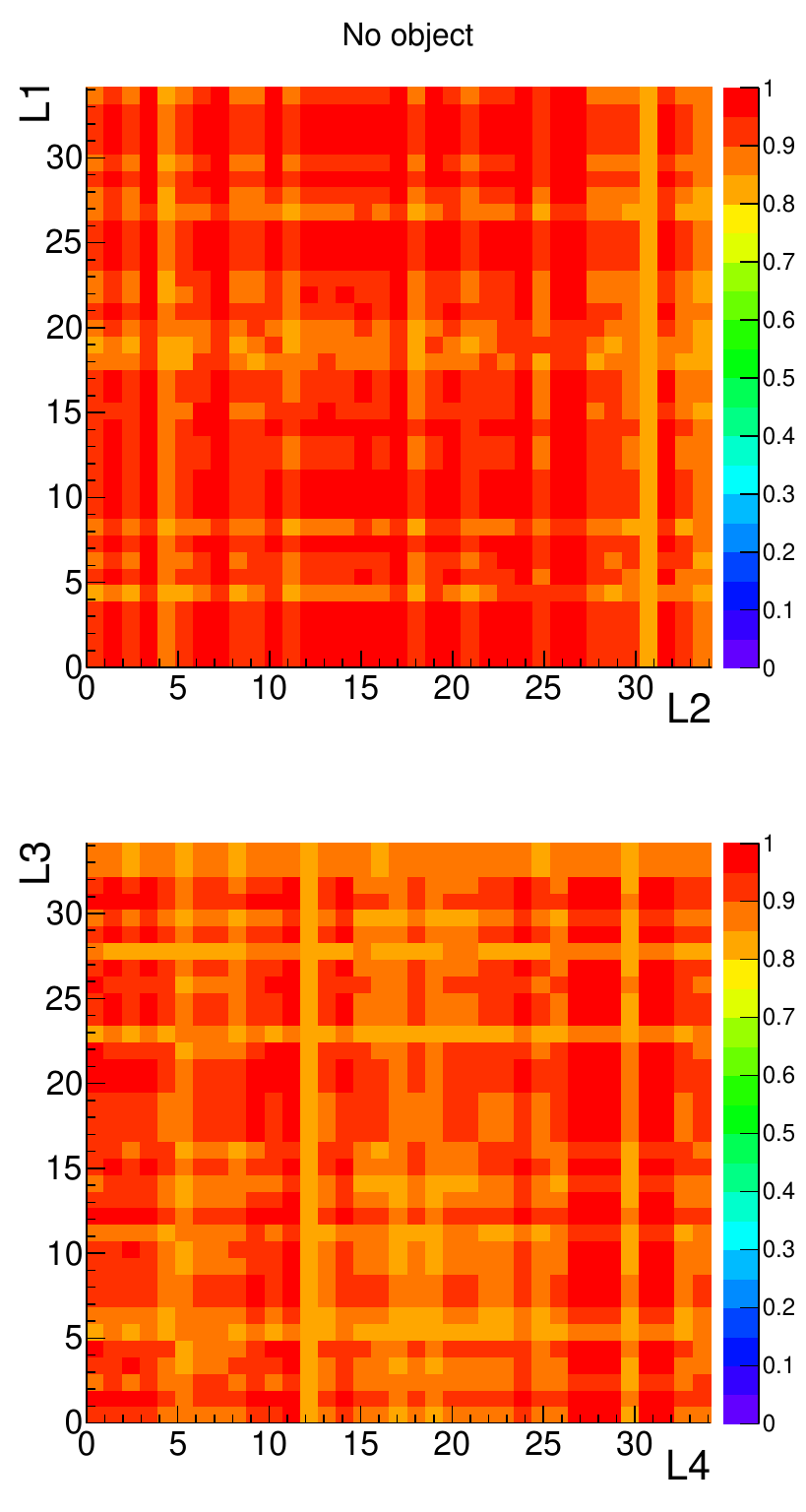}
\includegraphics[width=0.35\linewidth]{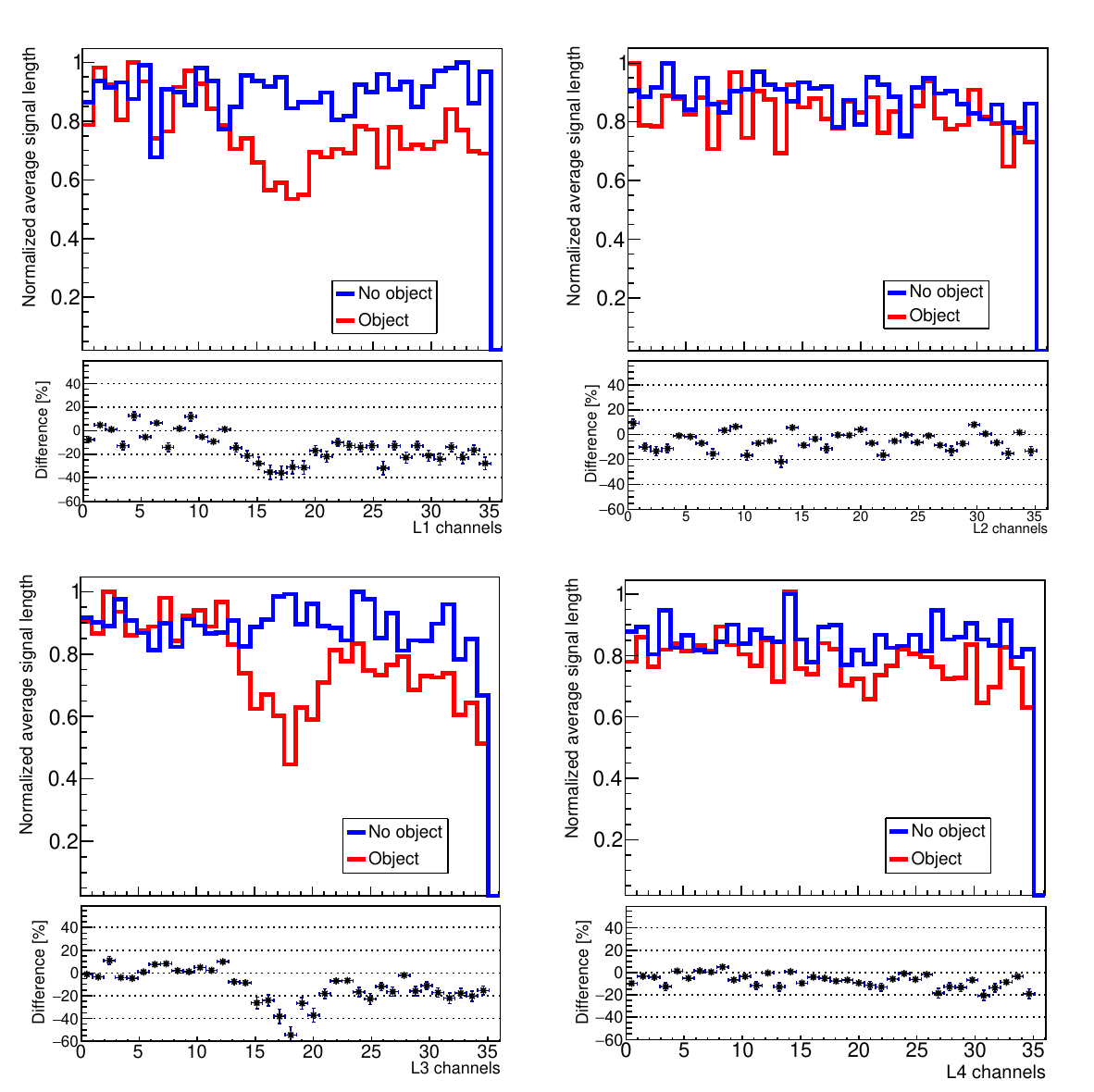}
\caption{{\em Top left:} Schematic of the detector plates showing the orientation of the detector bars (light-blue) and the lead object placed in front of the detector for the transmission measurement. The red arrow indicates the direction of the beam. {\em Top right:} A photo of the scanned lead object made of 10$\times$10$\times$5 cm$^3$ lead bricks in front of the van. The light-blue squares mark the position of the first detection module (L1-L2) inside the van.
{\em Bottom left:} Normalized 2D-maps of the measured average signal length, per bar, per pulse (30 pulses), for the front (L1-L2) and back (L3-L4) detection modules of the $\mu$36 detector, with or without a lead object placed in front of the detector. {\em Bottom right:} Normalized averaged signal lengths, per bar, per pulse, for each layer separately, with and without the object, and their relative differences.}
\label{fig:object_scan}
\end{figure}

For a series of 30 consecutive pulses, a 20$\times$25$\times$100 cm$^3$ lead object was placed at $\sim$30 cm in front of the first detection layer of the $\mu$36 detector. The object and the detector were located 23 m away from the converter. Figure \ref{fig:object_scan} shows normalized 2D-maps of the signal lengths for measurements with the object and without the object, for the two detection modules, layers L1-L2, closer to the object, and layers L3-L4 farther from the object. The bottom right plots show the signal length, per bar, for each individual layer.

First, it can be observed that there is a dip in the signal length measured in the L1 and L3 layers and no significant dip in L2 or L4. The 1-meter-long lead object shields the bars corresponding to channels 15 to 23 over their entire length for the L1 and L3 layers, whereas the L2 and L4 layers, with their bars placed perpendicular, are only partially shielded by the object, so that the bars are still exposed to the unaltered beam.

\begin{figure}[ht]
	\centering
    \includegraphics[width=0.78\linewidth]{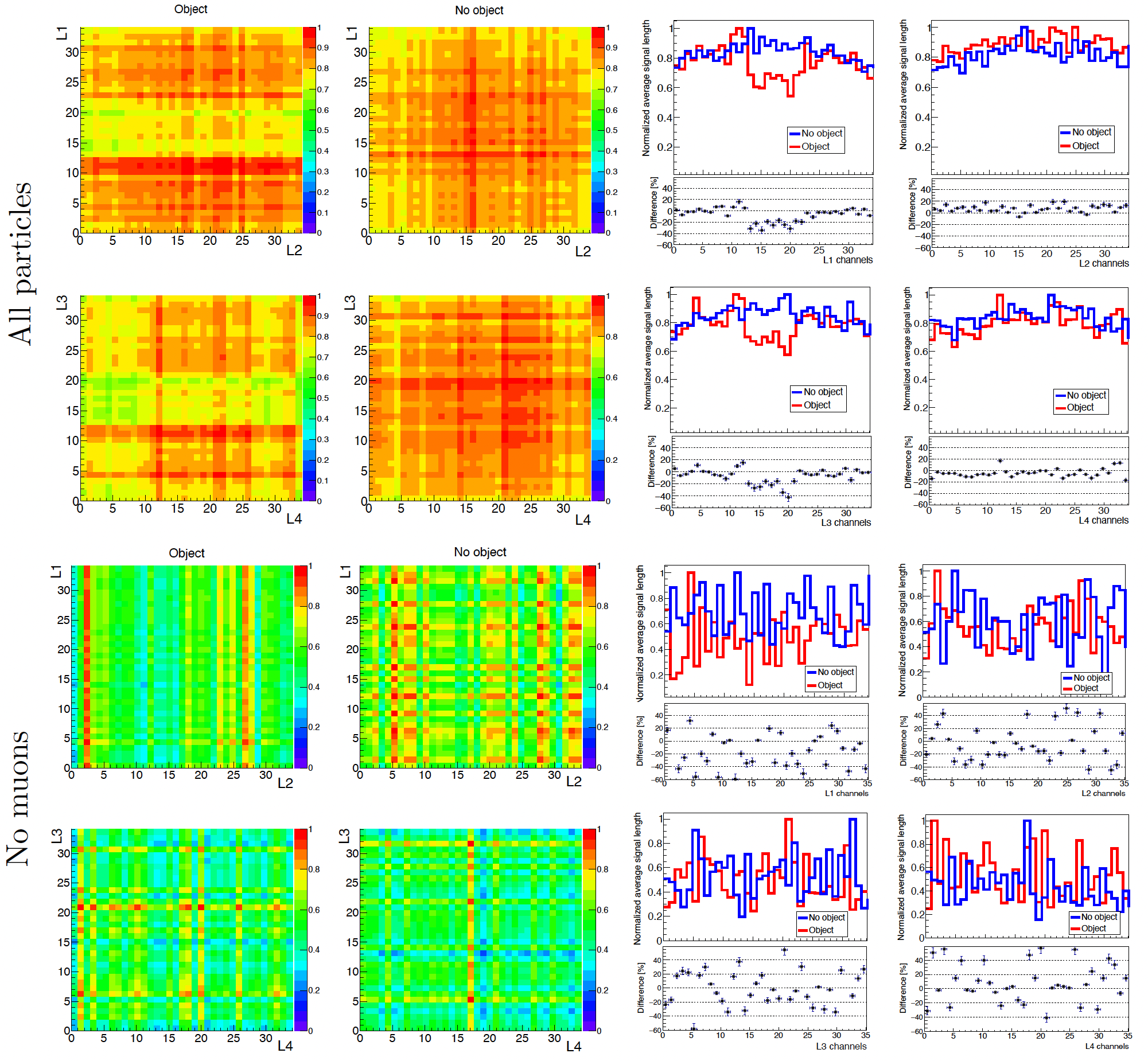}
	
	\caption{{\em Top left:} Normalized 2D-maps of the simulated average signal length, corrected for the time-over-threshold effect, for the front (L1-L2) and back (L3-L4) detection modules, per pulse, with or without a lead object placed in front of the detector, obtained from simulating the real pulses the $\mu$36 detector was exposed to during the measurements (30 pulses). {\em Top right:} Normalized averaged signal lengths, per bar, per pulse, for each layer, with and without the object, and their relative differences.
		{\em Bottom left:} Same simulated 2D maps as the top plots but including in the simulation only the background particles (gammas, neutrons, etc.), excluding the contribution of the muons. {\em Bottom right:} Normalized averaged signal lengths, per bar, per pulse, for each layer, with and without the object, and their relative differences for the background particles (gammas, neutrons, etc.).
	}
	\label{fig:object_scan_sims}
\end{figure}

The decrease in the measured signal length is observed over 9-10 bins. Since each bin corresponds to a scintillator bar 2.5 cm wide \cite{mu36:2025}, the reconstructed image corresponds to an object of 22.5-25 cm. This result, matching the dimensions of the real object (25 cm for the maximum thickness), in both L1 and L3 layers, confirms that the image was obtained with a directional, forward-oriented beam. 
Moreover, the average decrease in the signal length (26.17$\pm$7.31\% for L1, 26.19$\pm$9.20\% for L3) shows that enough particles had energies above the threshold required to pass through the object (254 MeV for a MIP muon perpendicular to the object). Such a beam is consistent with gamma-generated muons.
We have also simulated the detector response when the lead object is placed in front of the detector, starting from the same electron energy spectra of the 30 real pulses the detector was exposed to during the measurements. Figure \ref{fig:object_scan_sims} shows the same plots as Figure \ref{fig:object_scan} but obtained from GEANT4 simulations.

The dimensions of the object, reconstructed from the simulation, cover 10 bins in L1 and 9 bins in L3, consistent with the measurements and the dimensions of the object. The average decrease between simulations with the object or without the object is 33.62$\pm$13.34\% for L1 and 27.09$\pm$12.49\% for L3, 13\% higher than the average decrease for the measurements. 

While the intrinsic resolution of the detector is related to the bar size, 2.5 cm, the quality of the reconstructed image is influenced by the muon flux, background, or detector effects, such as noise, electronics response to high fluxes/light yield, or temperature dependence.
To investigate whether the observed transmission image could be obtained without muons, we have also simulated, for the same 30 measured pulses, the detector response to all laser-induced particles excluding the contribution of the muons. It can be seen in Figure \ref{fig:object_scan_sims}, bottom plot, that other particles, such as gammas and neutrons, cannot give the same image as the one generated by muons, likely due to their significant scattering and reduced energy deposit compared to the muons.

\section{The Time-over-Threshold correction}

\par The length of the signal observed in the $\mu$36 detector depends on the optical properties of the detector's components, such as light collection by the optical fiber, light absorption and re-emission efficiency of the wavelength-shifting fiber, SiPM sensitivity, optical contacts, as well as electronics effects such as amplifier clipping, and the non-linearity of the signal amplitude with incident number of photons caused by the Time-over-Threshold (ToT) \cite{mu36:2025,time_SiPM:2023}.

The ToT circuit induces a variable logarithmic relationship between the shape of the SiPM signal and the comparator threshold values in the circuitry \cite{Gonella:2015}.
For a SiPM signal, the amplitude is given by:
\begin{equation}
A(t) = \frac{A_0}{\tau} \left( e^{- \frac{t}{\tau}} \right)\\
\end{equation}
where $A_0$ is the integrated signal amplitude, estimated starting from the energy deposited in a bar, $E_{dep}$, which is further modified by optical factors.
\begin{equation}
A_0 = E_{dep} \cdot N_{photons} \cdot T_{fiber} \cdot PDE_{SiPM}
\end{equation}
$N_{photons}$ is the scintillator light yield ($\sim$10$^4$ photons/MeV for muons,  \cite{Luxium_datasheet}), $T_{fiber}$ is the light collection coefficient for the coupling used for this detector, which is $\sim$3\% according to the manufacturer \cite{fiber}, and $PDE_{SiPM}$ is the photon detection efficiency of the SiPM at the peak emission wavelength of the wavelength-shifting fiber of 494 nm, which amounts to 32 \%. The decay constant, $\tau$, is estimated at 90 ns 
and is a quantity dependent on the components of the detection chain, $\tau_{SiPM}$=80 ns
, $\tau_{fiber}$=2 ns, $\tau_{scint}$=2-3 ns, and $\tau_{electronics}$=5-6 ns.
The SiPM signal rise time, of the order 2-3 nanoseconds, was considered negligible.

Thus, the length of the signal above the ToT threshold of 0.4 V, calculated based on simulated energy deposit, is given by:
\begin{equation}
L_{corr} = \tau \cdot ln \left( \frac{A_0}{V_{th} \cdot \tau} \right)\\
\end{equation}

It must be noted that the ToT correction does not include other sources of noise/signal variation due to temperature fluctuations or SiPM signal nonlinearity with the number of incident photons. The detector was designed to record individual cosmic muons, for which the photon yield on the SiPM is rather low (a few tens of photons). Amplitude nonlinearity with the number of incident photons, at high photon fluxes, is not yet evaluated.

\section{Discussion and conclusions}\label{sec:conclusions}

Beams of muon pairs have been generated via the Bethe-Heitler process, by laser-wakefield-accelerated electrons interacting with a high-Z target. The muon beams were measured using detectors designed for muography applications.
The profile of the muon beam was measured at two distances away from the converter target, 20 and 29 meters, and showed a forward-peaked distribution with a FWHM of 4.51 m at 20 m and 6.35 m at 29 m, implying an angle of about 12.6$^{\circ}$.
Monte Carlo simulations performed with measured electron spectra as input confirmed that these observations are consistent with GeV muons (up to 6-7 GeV).
Propagation through the experimental set-up requires an energy threshold of 1.5 GeV (for survival probability $>$ 0) as the initial energy of the muon; therefore, the muon bunches reaching the detectors must have energies on the order of a few GeV.

Transmission measurements were performed to scan a lead object, and the observed image was successfully reconstructed from the measured signal length in individual bars of the detector. The intrinsic resolution is 2.5 cm, given by the bar size. The image was reconstructed using a signal dominated by muons, since simulations show that background particles (gamma photons, neutrons) cannot reproduce the result because of their pronounced attenuation and scattering compared to muons, and they generate only a flat background. The energy deposited by the muons is approximately one order of magnitude higher than the background.

To our knowledge, this represents the first time muon imaging has been performed with a laser-induced GeV muon signal prevailing overall. This was possible thanks to an appropriate shielding designed using Monte Carlo simulations that significantly reduced the background.
Future experimental campaigns envisage higher electron energies, coupled with the development of new detectors designed for high-flux, multi-particle discrimination, aiming for better beam characterization.

\begin{acknowledgments}
This work was supported by the Romanian National Authority for Research through the Nucleu Project No. PN 23 21 01 02, the ELI-NP Programme Component, contract RO-CH Nr. 1/ 30.10.2025, within the 2nd Swiss Contribution to Romania.

This work was partially supported by the ELI-NP project Phase II co-funded by the European Union through the European Regional Development Fund and by the Romanian Government through the Competitiveness Operational Programme. We also acknowledge the support of the Romanian Ministry of Research, Innovation and Digitization through the Nucleu Projects (Grant No. PN 23210105 and 19060105). Access to the ELI-NP facility is supported by the IOSIN funds for research infrastructures of national interest.

We also thank the ELI-RO contract $ELI-RO/RDI/2024-21$ LGS, which is funded by the Romanian Ministry of Research, Innovation, and Digitization.

The authors would like to thank A. Turturica and V. Vramulet (shielding), Vanessa Phung, Yinren Shou, and Vincent Lelasseux, S. B\u al\u a\c{s}cu\c{t}\u a and M. Gugiu, V. Nastasa (support in the experimental set-up).
\end{acknowledgments}

\bibliography{muon_imaging}

\end{document}